\documentclass[aps,pra,amsmath,amssymb,amsfonts,superscriptaddress,floatfix,nofootinbib]{revtex4} 
\pdfoutput=1
\usepackage{amsmath,amsthm,amsfonts,amssymb}
\usepackage{graphicx}

\usepackage{color}
\usepackage{hyperref}
\newcommand{\ab}{\kappa}
\newtheorem{theorem}{Theorem}
\newtheorem{conjecture}{Conjecture}
\newtheorem{corollary}{Corollary}

\usepackage{algorithm}
\usepackage[capitalize,nameinlink]{cleveref}

\newcommand{\be}{\begin{equation}}
\newcommand{\ee}{\end{equation}}

\newcommand{\cO}{{\cal O}}

\newcommand{\tg}{{\tilde \gamma}}

\newcommand{\expec}{\mathbb{E}}

\newcommand{\geqsos}{\underset{\rm SoS}{\geq}}
\begin{document}

\title{Field Theory and The Sum-of-Squares for Quantum Systems}
\author{Matthew B.~Hastings}
\begin{abstract}
This is a collection of various result and notes, addressing the sum-of-squares hierarchy for spin and fermion systems using some ideas from quantum field theory, including higher order perturbation theory, critical phenomena, nonlocal coupling in time, and auxiliary field Monte Carlo.
This paper should be seen as a sequel to Ref.~\cite{hastings2022optimizing} and Ref.~\cite{hastings2022perturbation}.
Additionally in this paper, we consider the difficulty of approximating the ground state energy of the Sachdev-Ye-Kitaev (SYK) model using other methods.  We provide limitations on the power of the Lanczos method, starting with a Gausian wavefunction, and on the power of a sum of Gaussian wavefunctions (in this case under an assumption).
\end{abstract}
\maketitle

\section{Introduction and Background}
The difficulty of simulating quantum systems on a classical computer has been recognized almost since the dawn of digital computers.  The world's first commercially available digital computer was the Ferranti Mark I.  Its successor was the Ferranti ``Mercury".  In 1964, Bonner and Fisher\cite{bonner1964linear} used a Ferranti Mercury to simulate a one-dimensional quantum spin chain with up to $11$ sites.  In this influential paper, they explicitly described the exponentially growing size of the Hilbert space (even using translation symmetry to reduce the size), and lamented that their ``relatively slow" computer restricted the available system sizes.  Of course, the importance of this exponential growth must have been understood well before that time, especially to anyone who attempted a hand calculation, but it is remarkable that some of the earliest computers were used to simulate quantum many-body problems.

In this regard, many of the simulation methods used in practice are attempts to find some polynomial time algorithm which gives reasonable answers at least in some regime.  For example, perturbation methods work at weak coupling, with a polynomial overhead at any given order of perturbation theory; DMRG and matrix product states work well in one-dimensional systems with low entanglement\cite{schollwock2011density}; and so on.

One particularly intriguing approach is based on the sum-of-squares hierarchy.  This method can give rigorous {\it lower} bounds on the ground state energy of a quantum system.  Any given level of the hierarchy takes a polynomial time, with the order of the polynomial increasing at higher levels of the hierarchy\footnote{This statement skips over some details.  At any given order of the hierarchy, one has a semidefinite progam of polynomial size, and commonly it is stated that such programs can be solved in polynomial time up to small additive error.  There can be some tricky hidden parameters in these claims (see \url{https://www.cs.cmu.edu/~odonnell/papers/sos-automatizability.pdf} , Ref.~\cite{OD17} ), but it seems that those issues cannot arise for the present problem.}.
See Refs.~\cite{GW95,CW04} for algorithms for classical systems, while for quantum systems with fermions the method is known as the reduced density matrix method (RDM)\cite{coleman1963structure,erdahl1978representability,
percus1978role,mazziotti2001uncertainty,Nakata_2001,Maz12,klyachko2006quantum}.  There is also a sum-of-squares hierarchy for qudit systems\cite{HM04,NPA08,DLTW08,PNA10}.

In this paper we give several results about the sum-of-squares.  These results are largely unrelated to each other, but there is a general theme that they have some relation to field theory methods.
In \cref{afqmc}, we discuss auxiliary field quantum Monte Carlo, showing a relation between optimal decompositions of the interaction (from the point of view of the sign problem) and a certain restricted version of the sum-of-squares method.
In \cref{ptferm}, we discuss the ability of sum-of-squares to reproduce perturbation theory for fermionic systems; we extend results of \cite{hastings2022perturbation}, which showed that second order perturbation theory could be reproduced by a fragment of degree-$6$ sum-of-squares but that it could not be reproduced by degree-$4$ sum-of-squares.  We show that for any given order of perturbation theory, there is some order of the sum-of-squares which reproduces it.  We do not, however, find the minimal order of sum-of-squares which can reproduce a given order of perturbation theory, and we leave this as an open question.
In \cref{critphen}, we discuss critical phenomena in the sum-of-squares framework, showing that in several cases the leading order sum-of-squares gives critical exponents which coincide with the large-$N$ $O(N)$ vector model.
In \cref{secnonlocal}, we consider methods related to the sum-of-squares applied to systems which have a \emph{nonlocal} interaction in time.
Finally, in \cref{sykground}, we consider the ability of various classical variational methods to approximate the ground state energy of the SYK model\cite{SY93,Kit15}.  While this section does not involve the sum-of-squares hierarchy directly, there is some relation because in Ref.~\cite{hastings2022optimizing} it was shown that sum-of-squares methods could certify one-sided bounds on the ground state energy within constant factors, with high probability.

\subsection{Background}
A review of the hierarchy is in Ref.~\cite{hastings2022optimizing}.  Some of the results extend results in Ref.~\cite{hastings2022perturbation}.  Since this paper is to some extent a sequel to those two papers, we will not give detailed definitions if they are explained there.

To very briefly sketch the idea of the hierarchy, one first chooses some set of operators, $\{O_a\}$.
In the so-called ``degree $2r$ sum-of-squares", the set of operators $\{O_a\}$ will be the set of monomials of degree at most $r$ in some operators, such as creation and annihilation operators for fermionic systems, or Pauli spin operators for some system of qubits.
Given this set of operators, one introduces a matrix, $M$, with matrix elements given by $$M_{ab}=\expec[O^\dagger_a O_b],$$
where $\expec[O^\dagger_a O_b]$ is a \emph{pseudo-expectation} value.
What does ``pseudo-expectation value" mean?  This means that we impose three conditions on $M_{ab}$.  First, we impose some linear relations, determined by the algebra of operators.  That is, if for some $\lambda_{ab}$ we have $\sum_{ab} \lambda_{ba} O^\dagger_a O_b=c$, for some scalar $c$, then we impose ${\rm Tr}(\lambda M)=c$.  Examples of these relations include things like $Z^2=1$ if $Z$ is a Pauli spin operator, or $\{\gamma_{a},\gamma_b\}=2\delta_{a,b}$ if $\gamma_a,\gamma_b$ are Majorana operators.
The second condition on $M$ is that $M$ is Hermitian.  The third condition is that $M$ is positive semi-definite.

Remark: because of these linear constraints, for degree-$2r$ sum-of-squares we can take the set $\{O_a\}$ to be a set which spans the same vector space of operators as the set of all monomials of degree at most $r$.  For example, with Maorana operators, we do not need both $\gamma_a \gamma_b$ and $\gamma_b \gamma_a$, but only need one of them.

Note that given any quantum density matrix $\rho$, the expectation value $\expec[O^\dagger_a O_b]={\rm Tr}(\rho O_a^\dagger O_b)$ is such a pseudo-expectation.  Such an $M$ is necessarily positive semi-definite because for any operator $O$, we have ${\rm Tr}(\rho O^\dagger O)\geq 0$.
Conversely, if the set of operators $\{O_a\}$ is complete in that every operator on the given Hilbert space is a linear combination of operators in the set, then every pseudo-expectation defines some quantum state $\rho$ such that $\expec[O^\dagger_a O_b]={\rm Tr}(\rho O_a^\dagger O_b)$, but if the set of operators $\{O_a\}$ includes, for example, only operators up to some given degree, then there may be pseudo-expectation values that do not correspond to any quantum density matrix $\rho$.

The sum-of-squares hierarchy defines a semi-definite program by minimizing the pseudo-expectation value of the Hamiltonian, subject to these constraints.  This is a semi-definite program, called the ``primal" problem.

It is standard that, given a primal semi-definite program which involves minimizing some quantity subject to constraints, there is a dual problem which involves maximizing some quantity subject to constraints, and the minimum of the primal is greater than or equal to the maximum of the dual.  In this particular case, the so-called ``duality gap" vanishes, and the minimum of the primal equals the maximum of the dual.
The dual problem has a particularly simple explanation in this case.  It is equivalent to:
given a Hamiltonian $H$, find a deomposition
$$H=\sum_\alpha \lambda_\alpha O_\alpha^\dagger O_\alpha+\lambda,$$ where $\lambda$ and $\lambda_\alpha$ are non-negative scalars and where the $O_\alpha$ are linear combinations of operators in the given set $\{O_a\}$ (e.g., polynomials of degree at most $r$).
For the optimal decomposition, $\lambda$ is equal to the ground state energy, and any such decomposition proves that the ground state energy is $\geq \lambda$.

Note that, for example, even if $H$ is a sum of terms which are degree at most $4$ in some variables, one may use $O_a$ which are polynomials of higher degree in such a decomposition, so that using the (anti-)commutation relations of the algebra one can show that the result is equal to $H$.
Of course, we may absorb $\lambda_\alpha$ into the definition of $O_\alpha$, rescaling $O_\alpha \rightarrow \sqrt{\lambda_\alpha} O_\alpha$ and writing $H=\sum_\alpha O_\alpha^\dagger O_\alpha+\lambda$, but sometimes it is more convenient to write it this way using $\lambda_\alpha$.

\subsection{Notation, and Conventions}
A remark on norms: given an un-normalized state (i.e., a positive semi-definite matrix), the norm we use, unless we say otherwise, is the $\ell_1$ norm.
So, given any state $\rho$, we say ``the projection of $\rho$ onto some subspace $S$ is $\leq \ldots$" to mean that the trace of the projection of $\rho$ onto $S$ is $\leq \ldots$

We use the notation $A \geqsos B$ to indicate that $A-B$ is a sum-of-squares, i.e., it is a sum of operators $O_a^\dagger O_a$.
If we are discussing a particular order of sum-of-squares, when we use $\geqsos$ we implicitly mean that the sum-of-squares is at most of that given order.

We use computer-science big-$O$ notation $o(n),\cO(n),\ldots$ throughout.
We use $n$ to indicate the number of fermionic modes or qubits, following \cite{hastings2022optimizing}.  We use $O(N)$ later to denote a particular orthogonal group.

We use $X_i,Y_i,Z_i$ to denote Pauli operators on a given qubit $i$ in a qubit system.
We use $\gamma_a$ to denote Majorana operators, $a\in\{1,\ldots,2n\}$.
We also use creation and annihilation operators $\psi^\dagger_a,\psi_b$ which obey canonical anti-commutation relations, with $a,b\in\{1,\ldots,n\}$.
Even for fermionic systems which might not obey particle-number conservation, we will see that there are some uses for creation and annihilation operators.

\section{Auxiliary Field Quantum Monte Carlo and the Importance of Commutators}
\label{afqmc}
The sum-of-square method above considers decomposition of a Hamiltonian $H$ as $H=\sum_\alpha O_\alpha^\dagger O_\alpha+\lambda$ for some scalar $\lambda$.  Here, the operator $O_\alpha$ may not be {\it normal}, meaning that the commutator $[O_\alpha^\dagger,O_\alpha]$ might not vanish.  Suppose indeed
$O_\alpha=A_\alpha+iB_\alpha$ for some Hermitian operators $A_\alpha,B_\alpha$, with $[A_\alpha,B_\alpha]\neq 0$.  Then
$$O_\alpha^\dagger O_\alpha=A_\alpha^\dagger A_\alpha+B_\alpha^\dagger B_\alpha+i [A_\alpha,B_\alpha].$$
The first two terms on the right-hand side are squares of Hermitian operators, but the commutator is not.

Indeed, if we restrict to sum-of-squares of Hermitian operators, so that we decompose $H=\sum_\alpha O_\alpha^2+\lambda$ for Hermitian $O_\alpha$, then the method becomes weaker than if we allow non-Hermitian $O_\alpha$.  For example, consider a system with two qubits with Hamiltonian
$$H=Z_1+Z_2+g X_1 X_2.$$
First consider a decomposition with non-Hermitian $O_\alpha$.  Let us consider the following, where $a,b,\lambda$ are some real scalars that we adjust (here we use the standard convention for Pauli matrices that $XY=iZ$):
$$H=\frac{1}{2}\Bigl(aX_1+ia^{-1}Y_1+bX_2\Bigr)\Bigl(aX_1-ia^{-1}Y_1+bX_2\Bigr)+1\leftrightarrow 2 +\lambda,$$
where $1\leftrightarrow 2$ means the same term as the first except with qubits $1$ and $2$ interchanged.
Then, we need to take
$$2ab=g,$$
and we have $-\lambda=a^2+(1+g^2/4)a^{-2}.$  Optimizing over $a$, we may obtain
$$\lambda=2\sqrt{1+g^2/4},$$ in agreement with the exact ground state.

Remark: this decomposition is almost the same as that used in \cite{hastings2022perturbation} to show that we may reproduce low order perturbation theory with the sum-of-squares, except that here we have allowed a more general decomposition with $a\neq 1$.  If we restrict to $a=1$, we reproduce low order perturbation theory.

Now suppose instead that we consider only a decomposition of $H$ as a sum of squares of Hermitian operators.  Again we consider the degree-$2$ sum-of-squares.  The dual formulation is in terms of a pseudo-expectation, $\expec[\cdot]$, as before.  However, now rather than requiring that the matrix of pseudo-expectation values $\expec[O_a^\dagger O_b]$ be Hermitian and positive semi-definite, we will take all operators $O_a$ defining that matrix to be Hermitian, and we will only require that the matrix be Hermitian and that the {\it symmetric} part of the matrix be positive semi-definite.
Taking the operators $O_a$ to be drawn from the set $\{1,X_1,Y_1,Z_1,X_2,Y_2,Z_2\}$, we consider the following pseudo-expectation.
The diagonal elements of the matrix of pseudo-expectation values are of course all equal to $+1$.
Let
$\expec[Z_1]=\expec[Z_2]=-1$; this of course implies that $\expec[1 Z_1]=\expec[Z_1 1]=\expec[1 Z_2]=\expec[Z_2 1]=-1$.
Assume $g>0$ and let
$\expec[X_1 X_2]=\expec[X_2 X_1]=-1.$
Finally, let all other matrix elements have vanishing real part; i.e., their contribution to the symmetric part of the matrix vanishes.
Note that the various linear constraints imposed on pseudo-expectation values by the Pauli commutation relations imply that
$\expec[X_1 Y_1]=-\expec[Y_1 X_1]=i\expec[Z_1]=-i$, i.e., the real part of that matrix element vanishes as required.  Similarly, $\expec[X_2 Y_2]=-\expec[Y_2 X_2]=-i$.

One may verify that this defines a matrix whose symmetric part is positive semi-definite but now we are only able to show that $H\geqsos -2-g$ at this order of sum-of-squares.

Given that this formulation of the sum-of-squares using only Hermitian term in the squares is weaker than the more general formulation, the reader may wonder why it is worth considering.  One answer is that it is worth discussing simply to emphasize why we want to use non-Hermitian operators in the squares.  However, there is also an interesting relation to auxiliary-field quantum Monte Carlo (AFQMC); see \cite{blankenbecler1981monte,sugiyama1986auxiliary} for original AFQMC papers, and see later work (too much to summarize here) for various methods of improving the sign problem.
Suppose we have a Hamiltonian $H$ for a fermion system which is a sum of quadratic and quartic terms.  Then, suppose we find a decomposition $H=Q+\sum_a Q_a^2+\lambda$ where the operators $Q,Q_a$ are Hermitian and are quadratic in the fermion fields; note that $Q$ is a quadratic term in the Hamiltonian while $Q_a^2$ includes quadratic and quartic terms. We choose $Q$ to be a sum-of-squares of linears in the fermion operators so $Q$ is positive semi-definite.  Then, we can implement an auxiliary-field Monte Carlo in imaginary time.  To do this, one may use Trotter-Suzuki to approximate the imaginary time evolution $\exp(-\beta H)$ by a product $$\exp(-\beta \lambda) \Bigl(\exp(-\tau Q) \prod_a \exp(-\tau Q_a^2) \Bigr)^{\beta/\tau},$$
where $\tau$ is a small timestep in imaginary time, and where the product $\prod_a \exp(-\tau Q_a^2)$
is taken in some arbitrary order.
Then, use a Hubbard-Stratonovich decoupling
$$\exp(-\tau Q_a^2) = (4\pi\tau)^{-1/2} \int \exp(i \phi Q_a) \exp\Bigl(-\frac{\phi^2}{4\tau}\Bigr) {\rm d}\phi.$$
This turns the evolution in imaginary time to the evolution under a quadratic Hamiltonian coupled to a fluctuating field.  In this case, there is a sign problem, but the magnitude of the sign problem depends on the difference between $-\lambda$ and the exact ground state energy.  That is, if the the ground state energy is $E_0$, then at large inverse temperature $\beta$, we have ${\rm tr}(\exp(-\beta \sum_a Q_a^2))\rightarrow \exp(-\beta(E_0-\lambda))$.  The operator  $\exp(i \phi Q_a)$ is unitary for any $\phi$, and so the decay 
$\exp(-\beta(E_0-\lambda))$ is due to a combination of the fluctuating sign due to averaging over different fluctuating auxiliary fields as well as any additional decay due to $\exp(-\tau Q)$.
To say it differently, the weight of any configuration is non-increasing as imaginary time increases (it may decrease due to the term $\exp(-\tau Q)$ but cannot increase); so, any decay in the average sign must lead to a decay in the total weight.
Thus, an optimal solution to the semi-definite program {\it may give an optimal decomposition for AFQMC} as we can lower bound the decay in the sign problem at long time by a constant times $\exp(-\beta (E_0-\lambda))$.

Of course, in many cases solutions using a real coupling to the auxiliary field, rather than imaginary, may lead to a better sign problem.

Some numerical experiments on quartic Hamiltonians show that there is a large loss in accuracy of the semi-definite program by using only Hermitian terms in the sum-of-square (for example, on some small molecules of 5-10 orbitals, changing from errors of $<10^{-4}$ Hartree using non-Heritian terms in the sum-of-squares to $\approx 0.2$ Hartree using only Hermitian terms).  However, even so the resulting error in the ground state energy ($0.2$ Hartree in this case) suggests that it might lead to a manageable sign problem.  When doing these calculations, we allowed the operators $Q_\alpha$ to be spin $0$ or spin $1$, and also allowed them to be particle-number-nonconserving; indeed, allowing that full generality was needed to obtain the optimal solution of the semidefinite program. 

Remark: this idea has some similarity of \cite{levy2021mitigating}, in that one finds some optimal way of writing a Hamiltonian to minimize a sign problem by a variational method.  Here the variational method is solving a semi-definite program, there the variational method involved maximizing an energy of a quantum Monte Carlo simulation.

For the rest of this paper, we consider the general case where the operators $O_\alpha$ may be non-Hermitian.

\section{Perturbation Theory for Fermionic Systems}
\label{ptferm}
Here we discuss the relationship between the sum-of-squares and perturbation theory.
Following \cite{hastings2022perturbation}, we consider a Hamiltonian
\be
\label{Hgen}
H=H_0+\epsilon \sum_{p=0}^4 H_{p,4-p},
\ee
where
\be
H_0=\sum_j E_j \psi^\dagger_j \psi_j,
\ee
where all $E_j$ are positive scalars, where $\epsilon$ is a small parameter controlling the perturbation theory, and where each term $H_{p,4-p}$ is a sum of products of $p$ creation operators and $4-p$ annihilation operators, with the term normal ordered so that the annihilation operators are to the right of the creation operators.  Thus, all terms $H_{p,4-p}$, except for $H_{4,0}$, annihilate the unperturbed ground state (i.e., when all number operators $n_i$ are equal to $0$).

For small $\epsilon$, there is a well-studied theory of perturbatively solving this Hamiltonian for the ground state energy as a function of $\epsilon$, which we denote $E_0(\epsilon)$.  Indeed, there are \emph{several} such perturbation methods, such as Rayleigh-Schrodinger, Brillouin-Wigner, and Green's function (diagrammatic) methods.  These methods all yield the same power series in the end, but may organize the computation differently.

We address two questions.  First, we consider a perturbative solution of the semi-definite program at a given order of the sum-of-squares.  Second, a related question, we consider whether the sum-of-squares at a given order can reproduce a given order of perturbation theory; here, we say that it reproduces a given order $k$ of perturbation theory if it proves a lower bound on the ground state energy which is at least $E_0(\epsilon)+o(\epsilon^{k})$.  Indeed, in all such cases where it does this we will find that it reproduces it up to error $\cO(\epsilon^{k+1})$.

Remark: of course, for physical system, the quantities $E_j$ in \cref{Hgen} may have either sign.  If they are negative, then the ground state of $H_0$ has some filled states.  However, by applying a particle-hole conjugation we can bring it into the form above with $E_j>0$.  This particle-hole conjugation also means that terms
$H_{p,4-p}$ may arise with $p\neq 2$ even if the original Hamiltonian conserves number.

\subsection{General Formalism for Perturbative Solution of Semidefinite Program and the Rank of the Reduced Density Matrix}
\label{genfpert}
We choose a basis for operators $O_a$ which are polynomials of degree at most $r$ in the creation and annihilation operators.  A suitable basis is to use normal ordered monomials, i.e., the annihilation operators are to the right of the creation operators.  We write such an operator as
$$O=\Psi^\dagger_{\vec u} \Psi_{\vec v},$$
where $\vec u,\vec v$ are bit strings of length $n$, where $n$ is the number of fermionic degrees of freedom.  Each operator $\Psi_{\vec v}$ is defined to be the product of $\psi_{a}$ for $a$ such that the $a$-th bit of $\vec v$ is nonzero, with the product taken in the order of increasing $a$.

When $\epsilon=0$, the exact ground state of $H$ is of course easy to find and one may calculate the expectation values of products of operators $O_a^\dagger O_b$ in this ground state.  Further, it is easy to show that any degree $2r$ sum-of-squares, for $r\geq 1$, reproduces all these expectations of monomials, for $O_a,O_b$ monomials of degree $d\leq r$.

In particular, the result for $\expec[O_a^\dagger O_b]$ is as follows.  Let 
$O_a=\Psi^\dagger_{\vec u_a} \Psi_{\vec v_a}$ and let
$O_b=\Psi^\dagger_{\vec u_b} \Psi_{\vec v_b}$.  Then $\expec[O_a^\dagger O_b]=0$ if $\vec v_b\neq 0$ or $\vec v_a\neq 0$.  If $\vec v_b=\vec v_a=0$, then $\expec[O_a^\dagger O_b]=\delta_{\vec u_a,\vec u_b}$, where
the $\delta$-function is a Kronecker delta-function.

With this choice of basis, and this solution of the semidefinite program for $\epsilon=0$, we may begin perturbation theory.  First, however, we remark on an interesting property regarding the rank of the matrix $M$ of pseudoexpectation values for a primal solution to the semidefinite program.

\subsubsection{Rank of $M$}
\label{rankM}
Note that at $\epsilon=0$ the matrix $M$ is diagonal, so the number of zero eigenvalues is simply equal to the number of zero entries on the diagonal.
The zero diagonal entries entries correspond to the case where $O_a=O_b=O$ with $O=\Psi^\dagger_{\vec u} \Psi_{\vec v}$ and with $\vec v$ nonzero.  
At $\epsilon=0$, the number of zero eigenvalues, keeping $O_a,O_b$ which are monomials of degree at most $r$, is equal to the number of choices of $\vec u,\vec v$ such that the total Hamming weight $|\vec u|+|\vec v|$ is $\leq r$ and such that $\vec v\neq 0$.  Thus, the number of zero eigenvalues equals
$$\sum_{s=1}^r \sum_{t=0}^{r-s} {n \choose s} {n\choose t}.$$

We emphasize that the matrix of \emph{pseudoexpectation} values $M$ obtained by solving the sum-of-squares hierarchy at any given order (of at least $2$) in the case $\epsilon=0$ reproduces the exact \emph{expectation} values in the ground state, and so in particular they have the same number of zero eigenvalues.

Now consider the following toy problem.  Take $n=4$ and let $$\sum_{i=1}^4 \psi^\dagger_i \psi_i+\epsilon(\psi^\dagger_1 \psi^\dagger_2 \psi^\dagger_3 \psi^\dagger_4+{\rm h.c.}),$$ where $+{\rm h.c.}$
means to add the Hermitian conjugate.

The exact ground state wavefunction can be written as a sum $\Psi_0(\epsilon)=a|0\rangle+b|4\rangle$ where $|0\rangle$ is the empty state (i.e., the state annihilated by all $\psi_i$) and $|4\rangle$ is the state with four particles (i.e., the state annihlated by all $\psi^\dagger_i$).  Indeed, the ground state energy $E_0$ is the lowest eigenvalue of the two-by-two matrix
$$\begin{pmatrix} 0 & \epsilon \\ \epsilon & 4\end{pmatrix},$$
and hence
$E_0=-2-\sqrt{4+\epsilon^2}.$

We now show that degree-$4$ sum-of-squares reproduces this, by writing
$H=E_0+\sum_\alpha \lambda_\alpha O_\alpha^\dagger O_\alpha,$ for some $O_a$.
We can guess an appropriate choice of $O_a$ by looking at the exact solution for the ground state wavefunction: we must have $O_a \Psi_0(\epsilon)=0$.
One choice is to pick a quadruple $i,j,k,l$ all distinct, with $i,j,k,l\in\{1,2,3,4\}$, and let
$O_\alpha=u \psi^\dagger_i \psi^\dagger_j + \psi_k \psi_l,$
where $u$ is a scalar.  
Without loss of generality, let us pick $i,j,k,l$ so that they give an even permutation of the sequence $1,2,3,4$ (if they are an odd permutation, then the sign of $u$ below is changed).
By inspecting the ground state wavefunction, we see that we need
$$u=\frac{\sqrt{4+\epsilon^2}-2}{\epsilon}.$$
Then, $\lambda_\alpha O_\alpha^\dagger O_\alpha=\lambda_\alpha u (\psi^\dagger_1 \psi^\dagger_2 \psi^\dagger_3 \psi^\dagger_4+{\rm h.c.})+\lambda n_i n_j + \lambda u^2 (1-n_k) (1-n_l),$
where $n_i=\psi^\dagger_i \psi_i$.
Summing over all choices of $i<j$, if we pick $$\lambda_\alpha=\lambda=\frac{\epsilon}{6u},$$ then
this gives the desired term $\epsilon (\psi^\dagger_1 \psi^\dagger_2 \psi^\dagger+3 \psi^\dagger_4+{\rm h.c.})$.

However, we do not yet have the correct term $\sum_i n_i$ in the Hamiltonian.
However, after some algebra, we find that
$$H=\sum_\alpha O_\alpha^\dagger O_\alpha+E_0(\epsilon)+c(\epsilon) \sum_{i<j} n_i (1-n_j),$$
for some non-negative scalar $c(\epsilon)$.  Further
$n_i (1-n_j)$ is a sum of squares, as $n_i (1-n_j)=O^\dagger O$ for $O=\psi_i \psi^\dagger_j$.
So, in this way we find the desired decomposition of $H$ as a sum of squares, giving the exact ground state energy at this order.

Now consider the rank of the matrix of pseudoexpectation values, $M$, for a solution of the semidefinite program, restricting to operators $O_a$ of degree at most $2$ in the fermionic operators.  Since degree-$4$ sum-of-squares reproduces the exact solution, we may consider the rank of the matrix of \emph{expectation} values in the true ground state.  One finds that the ground state is annihilated by any operator of the form $\psi^\dagger_i \psi_j$ for $i\neq j$, and there are $12$ such operators.  It also is annihilated by the operators
$u \psi^\dagger_i \psi^\dagger_j + \psi_k \psi_l$ and there are $6$ such operators.  Indeed, the number of zero eigenvalues is equal to $18$.
This compares with the case $\epsilon=0$ where we found that the number of zero eigenvalues equals
$\sum_{s=1}^2 \sum_{t=0}^{2-s} {4 \choose s} {4\choose t}=4+16+6=26.$

Thus, the rank of the matrix of pseudoexpectation values increases when $\epsilon$ becomes nonzero.  Indeed, even if consider the rank restricted to the submatrix where the operator $O_a$ has even fermion parity, then the number of zero eigenvalues at $\epsilon>0$ is equal to $18$ while the number at $\epsilon=0$ is equal to $22$ so even in that submatrix the rank has increased.

\subsubsection{General Formalism}
Now we develop a general formalism for a perturbative solution of the primal semidefinite
for a Hamiltonian of form \cref{Hgen} at a fixed order of the sum-of-squares.

The presentation here is not rigorous.  For example, we will ignore any questions of convergence of the series.

Assume we have a perturbative expansion $$M=M_0+\Delta,$$
where $M_0$ is the solution at $\epsilon=0$ and $\Delta$ is given by a series as
$\Delta=\epsilon M_1 + \epsilon^2M_2+\ldots.$

One simplification is that $M_0$, using the basis of operators $O_a$ above, is a projector.
Let $\Pi_0=1-M_0$.
We may perturbatively impose the requirement
$M\geq 0$, where the inequality is interpreted as meaning that $M$ is positive semi-definite.

To impose this requirement perturbatively, we need to ensure that the lowest eigenvalue of $M$ is $\geq 0$.
To do this, it is convenient to use Brillouin-Wigner perturbation theory.  This perturbation theory is simpler than Rayleigh-Schrodinger perturbation theory in the case of a degenerate ground state.  One complication that occurs in Brillouin-Wigner perturbation theory is that it involves denominators, $1/(E-E_i)$ where $E$ is the lowest eigenvalue of the \emph{perturbed} system and $E_i$ is some nonzero eigenvalue of the \emph{unperturbed} system; however we will see that these denominators simplify greatly.  
In our case, all these $E_i$ are equal to $1$.  The result of Brillouin-Wigner perturbation theory, is that there is some eigenvalue $E$ close to zero if
$$\Pi_0 \Bigl( \Delta + \Delta M_0 (E-1)^{-1} \Delta + \Delta M_0 (E-1)^{-1} \Delta M_0 (E-1)^{-1} \Delta + \ldots \Bigr) \Pi_0.$$
has an eigenvalue equal to $E$.

However, note that (assuming $\Delta$ is $\cO(\epsilon)$), reducing $E$ must increase the eigenvalues of this matrix since it increases the second order term $
\Delta M_0 (E-1)^{-1} \Delta$
and all higher terms $T$ in the series obey $T\leq \cO(\epsilon) \Delta M_0 (E-1)^{-1} \Delta$.
So, if there is a negative eigenvalue of this matrix for some $E<0$ then there is also a negative eigenvalue for $E=0$.

The result is that, perturbatively, positivity of $M$ is equivalent to the requirement that
\begin{eqnarray}
&&\Pi_0 \Bigl( \Delta - \Delta M_0 \Delta + \Delta M_0 \Delta M_0 \Delta + \ldots \Bigr) \Pi_0 
\\ \nonumber
&=& \Pi_0 \Delta \sum_{j=0}^\infty (-1)^j (M_0 \Delta)^j \Pi_0\\ \nonumber
&\geq & 0.
\end{eqnarray}

If we write $\Delta=\Delta_g+\Delta_e$ where $\Delta_g=\Pi_0 \Delta \Pi_0$, then this is equivalent to
$$\Delta_g+\Pi_0 \Delta_e \sum_{j=1}^\infty (-1)^j (M_0 \Delta_e)^j \Pi_0
\geq 0.$$
One obvious way to satisfy this is to have 
$\Delta_g=-\Pi_0 \Delta_e \sum_{j=1}^\infty (-1)^j (M_0 \Delta_e)^j \Pi_0$.
However, such a choice of $\Delta_g$ might not obey the linear relations imposed on $M$ by the canonical anticommutation relations.  Indeed, this is precisely why we considered the rank of $M$ in \cref{rankM}:
if we had $\Delta_g=-\Pi_0 \Delta_e \sum_{j=1}^\infty (-1)^j (M_0 \Delta_e)^j \Pi_0$ then the rank of $M$ would not change, but the rank does change in some cases.

It may be interesting to continue to develop this theory, to understand the perturbative solution of the sum-of-squares.  However, in the next section we turn to an alternative approach to show that the sum-of-squares can reproduce perturbation theory.

\subsection{Perturbation Theory and the Sum-of-Squares}
We now consider the question of reproducing a given order of perturbation theory using an appropriate order of the sum-of-squares.
For sufficiently small $\epsilon$, there is a power series in $\epsilon$ which defines a unitary $U(\epsilon)$ such that $\Psi_0(\epsilon)=U(\epsilon) \Psi_0(0),$ where $\Psi_0(\epsilon)$ is the ground state at given $\epsilon$ and $\Psi_0(0)$ is the unperturbed ground state.
We may prove this, for example, using exact\cite{osborne2007simulating,bravyi2011short}
quasi-adiabatic continuation\cite{hastings2004lieb} to construct a unitary describing the adiabatic evolution (for small enough $\epsilon$, the gap between ground and first excited state remains open, as needed for this method) of the ground state.  Alternatively, one may use higher-order Schrieffer-Wolff methods\cite{bravyi2011schrieffer}.

Let
$$\tilde \psi_i(\epsilon) \equiv U(\epsilon) \psi_i U(\epsilon)^\dagger,$$
for each $i$.
Since $\psi_i \Psi_0(0)=0$, we have
$$\tilde \psi_i(\epsilon) \Psi_0(\epsilon)=0.$$

The power series in $\epsilon$ for $U(\epsilon)$ defines a power series in $\epsilon$ for $\tilde \psi_i(\epsilon)$.  The term of order $\epsilon^k$
in this power series is a polynomial in the creation and annihilation operators $\psi^\dagger,\psi$ of degree at most $2k+1$.  
For brevity, let us simply say the term ``is a polynomial in $\psi,\psi^\dagger$".
The term of order $\epsilon^0$ is equal to $\psi_i$.

Similarly, we can define a power series in $\epsilon$ for $\psi_i$ where the term
term of order $\epsilon^k$
in this power series is a polynomial in operators $\tilde \psi^\dagger(\epsilon),\tilde \psi(\epsilon)$ of degree at most $2k+1$.
Again for brevity, let us simply say the term ``is a polynomial in $\tilde \psi,\tilde \psi^\dagger$".

Using this power series for $\psi$ in terms of $\tilde \psi$, we can write the Hamiltonian $H$ as a power series in $\epsilon$ where each term is a polynomial in $\tilde \psi,\tilde \psi^\dagger$.  
The term of order $\epsilon^0$ is equal to 
$$\sum_j E_j \tilde \psi_j(\epsilon)^\dagger \psi_j(\epsilon).$$
That is, it is the same as $H_0$ except with $\psi,\psi^\dagger$ replaced with $\tilde \psi,\tilde \psi^\dagger$.

Let $\tilde n_i(\epsilon)=\tilde \psi_i(\epsilon)^\dagger \tilde \psi_i(\epsilon).$

Note that we have canonical anti-commutation relations also for $\tilde\psi,\tilde\psi^\dagger$, i.e., $\{\tilde\psi_j(\epsilon),\tilde \psi_k(\epsilon)\}=\{\tilde \psi_j(\epsilon)^\dagger,\tilde \psi_k(\epsilon)^\dagger\}=0$ and $\{\tilde\psi_j(\epsilon)^\dagger,\tilde \psi_k(\epsilon)\}=\delta_{j,k}.$
So, we may take the representation of $H$ in terms of 
$\tilde \psi,\tilde \psi^\dagger$ and normal order the terms, using these anti-commutation relations.
This then expresses
$H=\sum_j E_j \tilde n_j(\epsilon)+V+\lambda,$
where $V$ is $\cO(\epsilon)$ and is a sum of normal ordered terms and $\lambda$ is a scalar.

Indeed, since the terms $V$ are normal ordered, then for $\epsilon$ small enough that
$\Psi_0(\epsilon)=U(\epsilon) \Psi_0(0)$, we have
$\lambda=E_0(\epsilon)$, i.e., it is the ground state energy at the given $\epsilon$.

Moreover, by lemma 1 of Ref.~\cite{hastings2022perturbation}, each normal ordered term in $V$ of degree $d$ is equal to some linear combination of $\tilde n_j$ plus a sum-of-squares of terms, each term being a polynomial in $\tilde \psi,\tilde \psi^\dagger$ degree at most $d/2$.  The sum, over all terms in $V$ of all these linear combinations $\tilde n_j$ can be written as some $\sum_j \delta_j(\epsilon) \tilde n_j$.
So, we express
$H\geqsos \sum_j (E_j+\delta_j(\delta)) \tilde n_j(\epsilon)+E_0(\epsilon).$  Since $\delta_j(\epsilon)$ is $\cO(\epsilon)$, for small enough $\epsilon$ we have $E_j+\delta_j(\epsilon)>0$ for all $j$ so 
$\sum_j (E_j+\delta_j(\delta)) \tilde n_j(\epsilon)$ is a sum of squares.

Now let us show that for any given order of perturbation theory, some finite order of the sum-of-squares can reproduce the results at that order.

We have, as explained above, a sum-of-squares proof $H\geqsos \sum_j (E_j+\delta_j(\delta)) \tilde n_j(\epsilon)+E_0(\epsilon).$  Indeed, this means that $H=\sum_\alpha O_\alpha^\dagger O_\alpha+E_0(\epsilon),$ where each $O_\alpha$ is a polynomial in operators $\tilde \psi,\tilde \psi^\dagger$ and where we have fixed $\lambda_\alpha=1$ and so omitted $\lambda_\alpha$.  

Further, we claim that each $O_\alpha$ has either even or odd fermion parity, i.e., it is a polynomial with only terms of even degree in $\tilde \psi,\tilde \psi^\dagger$ or with only odd degree in $\tilde \psi,\tilde \psi^\dagger$, as the Hamiltonian only has terms of even degree (this is a special case of a more general result on symmetries of a Hamiltonian discussed in \cref{critphen}).

We claim the coefficient of a term of degree-$q$ in $\tilde \psi,\tilde \psi^\dagger$ in $O_\alpha$ is 
$\cO(\epsilon^{(q-1)/2})$.  Thus, a cubic term is of order $\epsilon$, while a quadratic term is of order $\sqrt{\epsilon}$.  Intuitively, this makes sense: the square of the quadratic term is a quartic term, and the quartic term in the Hamiltonian is of order $\epsilon$.  However, we may prove that this must be the case in general as follows: the term of degree-$2$ in $H$ is $\cO(1)$ and the term of degree-$4$ in $H$ is $\cO(\epsilon)$, and so using the series for $\psi,\psi^\dagger$ in terms of $\tilde \psi,\tilde \psi^\dagger$ expresses the Hamiltonian $H$ in terms of $\tilde \psi,\tilde \psi^\dagger$ where each term of degree $d$ is of order $\cO(\epsilon^{(d-2)/2})$.  Normal ordering does not change this: we get a normal ordered Hamiltonian in terms of $\tilde \psi,\tilde \psi^\dagger$, where again each term of degree $d$ is of order $\cO(\epsilon^{(d-2)/2})$, as normal ordering can only reduce the degree.  Finally, 
when using lemma 1 of \cite{hastings2022perturbation} to that show each normal ordered term in $V$ of degree $d$ is equal to some linear combination of $\tilde n_j$ plus a sum-of-squares of terms, each term having degree at most $d/2$, the resulting sum-of-squares has the desired property that the coefficient of a term
of degree-$q$ is $\cO(\epsilon^{(q-1)/2})$.

Next, one may re-express each $O_\alpha$ as a series in $\psi,\psi^\dagger$, 
using the series for $\tilde \psi,\tilde \psi^\dagger$ in terms of $\psi,\psi^\dagger$.
Then, again we have a similar result: the coefficient of a term of degree $r$ in $\tilde \psi,\tilde \psi^\dagger$ in $O_\alpha$ is 
$\cO(\epsilon^{(r-1)/2})$.
Suppose we truncate the series for each $O_\alpha$ in $\psi,\psi^\dagger$ at degree $r$ for some $r$.
For $r=2k+1$, this truncation of $O_\alpha$ is correct up to error $\cO(\epsilon^{k+1})$.
To clarify what we mean by ``correct up to error $\ldots$", since the quantity we are talking about is an operator rather than a number, we mean simply that it is a polynomial whose coefficients are of the given order.

For example, for $r=3$, we can express the leading term $\tilde \psi^\dagger_i \tilde \psi_i$ up to
error $\cO(\epsilon^2)$, because we include the terms of degree $3$ but not the terms of degree $5$.

Call this truncation $O_\alpha^{\rm trunc}$.
Then, $H=\sum_\alpha \lambda_\alpha (O_\alpha^{\rm trunc})^\dagger O_\alpha^{\rm trunc}+E_0(\epsilon)+\delta,$ where $\delta$ is the ``error" in making this truncation.  The term $\delta$ is, by construction, $\cO(\epsilon^{k+1})$.  Further, $\delta$ is of degree at most $2r$ by construction, assuming $2r\geq 4$, as then every term in $H$ and in $\sum_\alpha \lambda_\alpha (O_\alpha^{\rm trunc})^\dagger O_\alpha^{\rm trunc}$ is at most
degree $2r$.  
So, we may show that $\delta\geqsos -\cO(\epsilon^{k+1})$ by a degree-$2r$ sum-of-squares proof\footnote{This is a trivial proof.  There is a sum-of-squares proof that any monomial in $\psi,\psi^\dagger$, with coefficient equal to $1$, is $\geq -1$ and $\leq +1$.}.

So, this gives a degree-$2(2k+1)$ sum-of-squares proof that $H\geqsos E_0(\epsilon)-\cO(\epsilon^{k+1})$,
i.e., a degree-$4k+2$ proof.
So, as claimed, for any desired order of perturbation theory, there is some order of the sum-of-squares that reproduces it, i.e., $k$-th order perturbation theory is reproduced by degree $4k+2$ sum-of-squares.
However, we can see that this result is not optimal.  In \cite{hastings2022perturbation}, it was shown that degree-$6$ sum-of-squares reproduces second order perturbation theory.

We leave it as an open question to determine what the minimal order of sum-of-squares is to reproduce a given order of perturbation theory.  In \cite{hastings2022perturbation} it was proven that second order perturbation theory is reproduced by degree-$6$ sum-of-squares and not by degree-$4$.

\section{Critical Phenomena}
\label{critphen}
In this section, we apply leading order sum-of-squares to various models with a quantum critical point.  Interestingly, we see exponents that coincide with the large-$N$ vector model (explained below) in a variety of cases.  We see this both when the model is a vector model, and when it is a transverse field Ising model.  The next-to-leading-ordered sum-of-squares treatment of this critical phenomena may be very complicated.  

\subsection{Large $N$ Vector Model---Relation To Sum-of-Squares}
The $O(N)$ vector model is a model studied in quantum field theory (see chapter 8 of \cite{polyakov1987gauge} or section 4 of \cite{zinn1998vector}).
(The notation $N$ in the vector model should not be confused with our use of $n$ elsewhere for the number of degrees of freedom.)
  We can give a Hamiltonian formulaton of this model on a lattice as follows.  Consider a lattice of sites labelled by integers $i,j,\ldots$; for example, consider a cubic lattice in $d$ spatial dimensions.  On each site $j$, there are $N$ continuous degrees of freedom, for some integer $N\geq 1$.  
To describe these degrees of freedom, we introduce operators $q_j^\mu$ and $p_j^\mu$, where $\mu\in\{1,\ldots,N\}$ indexes the different degrees of freedom on the given site $j$. These operators obey the canonical commutation relations $[q_j^\mu,p_j^\nu]=i \delta_{\mu,\nu}$, where $\delta_{\mu,\nu}$ is the Kronecker $\delta$-function (fixing $\hbar=1$).  We use $\vec q_j$ to denote a vector with components $q_j^{\mu}$ and similarly $\vec p_j$ denote a vector with components $\vec p_j^\mu$.

The Hamiltonian is then
\be
H=J \sum_{<i,j>} \vec q_i \cdot \vec q_j+\frac{1}{2} \sum_i (\vec p_i)^2 + \frac{V}{2} \sum_i \Bigl(\frac{(\vec q_i)^2}{N}-1\Bigr)^2,
\ee
where the notation $\sum_{<i,j>}$ denotes the sum over nearest neighbor $i$ and $j$ and where
$(\vec p)^2\equiv \vec p \cdot \vec p$.
Taking $V$ large forces $(\vec q_i)^2$ to be close to $1$, so that the vector $\vec q_j$ is constrained to given length.  

At large $N$, it is possible to solve this model using a saddle point method; one decouples the quartic interaction in $H$ using a Hubbard-Stratonovich transformation, introducing an auxiliary field (which becomes our quantity $\ab$ below), and then takes a saddle point in the integral over the auxiliary field (the saddle point is accurate for large $N$, and it gives the self-consistent equation below).  This solution does not require $V$ to be large.  The resulting solution gives an interesting solvable model that displays a phase transition with non-mean-field behavior in three dimensions.
At small $J$, the different sites are approximately decoupled, showing paramagnetic behavior, but for $J$ larger than some critical $J_c$, long-range ferromagnetic order sets in.

The solution of this large $N$ model can be summarized as follows.  For simplicity, we consider the model on a cubic lattice, which simplifies the construction due to translational invariance.
Then, there is some scalar $\ab$\footnote{Without translation invariance, it becomes necessary to introduce a different mass for each site.}.  Then, we consider the Hamiltonian
$
H_{\rm Gaussian}\equiv\sum_{<i,j>} \vec q_i \cdot \vec q_j+\frac{\ab}{2} \sum_i (\vec q_i)^2+\frac{1}{2}\sum_i (\vec p_i)^2.
$
This Hamiltonian describes coupled harmonic oscillators and can be readily solved using creation and annihilation operators.  To do this, one applies a discrete Fourier transform to go to normal modes.

The ground state of the Hamiltonian has some given expectation value $\expec[\frac{1}{N}(\vec q_j)^2]$.  Then, $\ab$ is chosen to satisfy a self-consistent equation
\be
\label{m}
\frac{\ab}{2}=V\Bigl(\expec\Bigl[\frac{1}{N}(\vec q_j)^2\Bigr]-1\Bigr).
\ee

As one approaches $J_c$, the quantity $\ab$ displays interesting critical behavior.  As $J$ approaches $J_c$ from below, in the infinite lattice size limit, $\ab$ tends to some quantity such that, at that quantity, the energy of the lowest normal mode of $H_{\rm Gaussian}$ is equal to zero; this lowest normal mode is the one at wavevector equal to $0$.  However, for any finite lattice size, $\ab$ never reaches that quantity.  For $J>J_c$, at finite lattice size, a non-negligible contribution to $\expec[\frac{1}{N}(\vec q_j)^2]$ comes from the lowest normal mode.

Following this very brief review, it is interesting to see that the sum-of-squares method reproduces the same self-consistent solution for $\ab$ even at $N=1$.
For the case $N=1$, we will drop the vector notation, and simply use operators $q_j$ and $p_j$.

We use the following fact:
for any real scalar $s$, the function $(x-1)^2$ obeys $(x-1)^2\geqsos (s-1)^2+2(x-s)(s-1)$; note, the right-hand side are the zeroth and first order terms of a Taylor expansion of $(x-1)^2$ around $x=s$.  We have $(s-1)^2+2(x-s)(s-1)=1-s^2+2x(s-1)$.

So, for any $s$,
\be
\label{Vsos}
\frac{V}{2}(q_j^2-1)^2\geqsos \frac{V}{2}(1-s^2)+V q_j^2(s-1).
\ee
So,
for any $s$,
we have
$$H\geqsos H_{\rm Gaussian}+\sum_j \frac{V}{2} (1-s^2),$$
where we take
$$\frac{\ab}{2}=V(s-1).$$

The ground state energy of $H_{\rm Gaussian}$ can be calculated exactly using degree-$2$ sum-of-squares; indeed, this is precisely the usual calculation using creation and annihilation operators.
We now vary over $s$ to get the tightest lower bound on the ground state energy.
The derivative of the ground state energy of $H_{\rm Gaussian}$ with respect to $s$ is
$$\frac{\partial \ab}{\partial s}\sum_j \frac{1}{2} \expec[q_j^2]=V \expec[q_j^2].$$
Hence, the condition for a maximum is
$V \expec[q_j^2]=Vs$, so
$$s=\expec[q_j^2].$$
Using the given $s$ and $\ab$, we see that this is the same self-consistent equation as \cref{m} for $N=1$.

The reader might note that we are working in a slightly non-standard form of the sum-of-squares.  \cref{Vsos} is an inequality in the degree-$4$ sum-of-squares, but otherwise we use only the degree-$2$ sum-of-squares to solve $H_{\rm Gaussian}$.  That is, we use part of the degree-$4$ sum-of-squares (indeed, we must, since the Hamiltonian is degree-$4$), but otherwise we use only the degree-$2$ sum-of-squares.  
We claim, but leave to the reader to show, that we have found the optimal dual solution if we consider degree-$2$ sum-of-squares as well as sums of, for each $j$, squares of polynomials of at most degree $2$ in $q_j$.  So, it is an interesting question how using the full power of degree-$4$ sum-of-squares would change the solution.

To emphasize why it is slightly surprising that sum-of-squares gives the same solution at $N=1$, consider an alternative approximate method of solving the case $N=1$.  This alternative method gives a variational upper bound on the ground state energy.  We use the ground state of Hamiltonian $H_{\rm Gaussian}$ as a variational state, and compute the expectation value of Hamiltonian $H$ in this state, minimizing over $\ab$.  If one works out the details, one will find a different self-consistent equation for $\ab$.  Briefly stated, the reason is that sum-of-squares effectively uses the inequality
$\expec[(q_j^2)^2]\geq \expec[q_j^2]^2$, while in the Gaussian state
we have $\expec[(q_j^2)^2]=3\expec[q_j^2]^2$.

\subsection{Transverse Field Ising Model at Leading Order}
We now turn to the transverse field Ising model.

\subsubsection{Mean Field Theory}
We begin with a treatment of a Hamiltonian appropriate for mean-field theory.  Consider
\be
H=-\frac{1}{2n}\sum_{i,j} Z_i Z_j-h\sum_i X_i,
\ee
where there are $n$ qubits, with corresponding Pauli operators $X_i$ and $Z_i$.  At small $h$, we expect a ferromagnetic phase in the limit of large $n$, while at large $h$ we expect a paramagnet.  The factor of $1/2$ is to avoid double counting.

We can solve this Hamiltonian in a mean-field approximation.  Take a product state for the spins, where each spin has $\langle X_i \rangle=\cos(\theta)$ and $\langle Z_i \rangle=\sin(\theta)$ for some angle $\theta$.
Then the expectation value of the energy is (up to $\cO(1)$ corrections for the case $i=j$ in the sum)
$$-\frac{n}{2} \sin^2(\theta)-n h \cos(\theta).$$
For $h\geq 1$, the minimum is at $\theta=0$, with energy $-n h$.  For $h\leq 1$, the minimum is at $\cos(\theta)=h$, with energy
$$-\frac{n}{2}(1-h^2)-nh^2=-\frac{n}{2}(1+h^2).$$

Now we consider a sum-of-squares treatment of the problem.
We pause first for a useful result.  Suppose we have some symmetry group of a Hamiltonian $H$, i.e., we have a group homomorphism $\pi$ from some group $G$ to the group of unitaries, such that the unitaries in the image commute with $H$.
Further, suppose that, acting by conjugation, these unitaries do not increase the degree of a monomial, i.e., given an operator $O$ of given degree in some operators (e.g., the Pauli operators) and given a $g\in G$, the operator $\pi(g) O \pi(g)^\dagger$ has the same degree as $O$.
  Then we claim that given a sum-of-squares representation as
$H=\sum_\alpha O_\alpha^\dagger O_\alpha+\lambda$, we can find a sum-of-squares representation
$H=\sum_\alpha Q_\alpha^\dagger Q_\alpha+\lambda$ of the same degree, where each $Q$ has the property that it maps under conjugation by $\pi(g)$ according to some irreducible representation of $G$.  To see this, suppose $O_\alpha$ is a sum of operators which transform by conjugation under inequivalent irreducible representations, i.e., $O_\alpha=\sum_r O_{\alpha,r}$ where $r$ labels irreducible representations and $O_{\alpha,r}$ transforms by conjugation according to representation $r$.  Then,
\be
\label{crossterm}
O_\alpha^\dagger O_\alpha=\sum_{r,r'} O_{\alpha,r}^\dagger O_{\alpha,r'}.
\ee  By assumption that $g$ is a symmetry of $H$, for any $g\in G$ we may make the replacement $O_\alpha\rightarrow \pi(g) O_\alpha \pi(g)^\dagger$ for every $\alpha$, and this gives another sum-of-squares representation of $H$.  By summing over this replacement for various choices of $g$, we may remove any ``cross-terms" in \cref{crossterm}, i.e., remove those terms with $r\neq r'$.  

In this case, the symmetry group that we will use is the symmetry under spin flip, meaning $\prod_i X_i$, as well as a symmetry under cyclic permutation of the spins $1\rightarrow 2\rightarrow 3\ldots$  Indeed, we have a full spin permutation symmetry but we will not need that.

We will use degree-2 sum-of-squares, so we search for a representation
$H=\sum_\alpha O_\alpha^\dagger O_\alpha$ where each $O_\alpha$ is degree $1$.
By the above general result, we may assume each $O_\alpha$ is either even or odd under spin flip (i.e., stays the same or changes sign under spin flip).

We will make a few choices, where we say it ``suffices" to consider only certain things; one may verify that these indeed give the optimal decomposition.  Further, since as we will see the result agrees with
mean-field up to $\cO(1)$ corrections in the energy, it implies that these choices do give the correct energy up to $\cO(1)$ corrections.
First, it suffices to consider only odd terms.  So, $\alpha$ labels different irreps under cyclic permutation.  Also, it suffices to take only one term for each irrep.

So, we represent
\be
H=\sum_k O(p)^\dagger O(p) + \lambda,
\ee
where $p$ ranges over $0,1,\ldots,n-1$
and
$$
O(p)=\frac{1}{\sqrt{n}}\sum_p \exp(i j p) \Bigl(a(p) Z_j + i b(p) Y_j\Bigr),
$$
where it suffices to take $a(p),b(p)$ as real scalars.  Here $p$ plays the role of a ``momentum", i.e., labeling different Fourier modes

Due to the symmetry under arbitrary permutation of sites, we may assume that all $a(p)$ are the same for $p\neq 0$ and similarly for $b(p)$.  So, we set $a(p)=a$ for $p=0$ and $a(p)=a'$ for $p\neq 0$ and $b(p)=b$ for $k=0$ and $b(p)=b'$ for $b\neq 0$.
Since the Hamiltonian has no $Y_i Y_j$ terms in it, indeed we must have $b=b'$.
Then, to obtain the correct $Z_i Z_j$ term in $H$
we need
$$
a^2=a'^2-\frac{1}{2}.
$$
To obtain the correct $X_i$ term in $h$ we need
$$ab+(n-1)a'b=hn/2.$$

From this we get
$$-\lambda=b^2n+(a')^2 (n-1)+a^2,$$ and we wish to minimize $-\lambda$ (i.e., maximize $\lambda$).
Let us solve this for large $n$.
Then we approximate $a'b=h/2$ and we wish to minimize $n(b^2+a'^2)$ subject to $a'^2\geq 1/2$ since $a^2\geq 0$.
For $h\geq 1$, the minimum is at $b=\sqrt{h/2}$ with
$\lambda=-hn+o(n)$; recall that $o(n)$ denotes a term asymptotically smaller than $n$.

For $h\leq 1$, we have $a'=1/\sqrt{2}$ and so $b=h/\sqrt{2}$ and so
$\lambda=-\frac{n}{2}(1+h^2)+o(n)$.
So, the sum-of-squares result matches the variational result, up to corrections which are subleading in $n$.

\subsubsection{Three Dimensions}
We now turn to the three-dimensional Ising model on a cubic lattice with $n$ sites.  We label sites by triples of integers, using a vector notation such as $\vec j$ to label a site.
We let
\be
H=-\frac{1}{2}\sum_{<\vec j,\vec k>} Z_{\vec j} Z_{\vec k} + h \sum_{\vec j} X_{\vec j},
\ee
where the sum $\sum_{<\vec j,\vec k>}$ is over nearest neighbor $\vec j,\vec k$.

The Hamiltonian has a spin flip symmetry as before.  It also has a symmetry under translation by $1$ site in any of three orthogonal directions.  We will label Fourier modes by vectors $\vec p$, e.g., given an $L$-by-$L$-by-$L$ cube with $L^3=n$, we have $\vec p=(p_x,p_y,p_z)$ where $p_x,p_y,p_z$ are integer multiplies of $2\pi/L$.

As in the previous subsection, we consider a decomposition of $H$ as a sum-of-squares as
\be
H=\sum_{\vec p} O(\vec p)^\dagger O(\vec p)+\lambda,
\ee
where
$$O(\vec p)=a(\vec p) Z(\vec p) + i b(\vec p) Y(\vec p),$$
where $a(\vec p),b(\vec p)$ are real scalars
and where
$Z(\vec p)=n^{-1/2} \sum_{\vec j} \exp(i \vec p \cdot \vec j) Z_{\vec j}$
and
$Z(\vec p)=n^{-1/2} \sum_{\vec j} \exp(i \vec p \cdot \vec j) X_{\vec j}$.
The reader may verify that this is an optimal decomposition at this order; we omit the proof.

As in the mean-field case, since the Hamiltonian has no $YY$ terms, $b(\vec p)$ must be independent of $\vec p$.  So, we write $b(\vec p)=b$ for some $b$.

Given the $ZZ$ terms in $H$, we must have
$$a(\vec j)^2=-\cos(j_x)-\cos(j_y)-\cos(j_z)+c,$$
for some scalar $c$.
To get the correct $X$ term in $H$, we need
\be
\label{sumneed}
\frac{1}{n}\sum_{\vec p} a(\vec p) b = h.
\ee

We have $\lambda=-n(b^2+c)$, so we wish to minimize $b^2+c$.

We will be concerned only with what the criticial behavior is, so we will make various approximations.
In an integral approximation to \cref{sumneed}, we need
\be
\label{consistency}
\int  a(\vec p) \,\frac{{\rm d}^2 p}{(2\pi)^2}=\frac{h}{b}.
\ee
We will work in an approximation: we consider only $\vec p$ close to $0$, and we expand
$$a(\vec p)^2\approx -3+\frac{1}{2}\vec p^2+c.$$
Let us write $m^2=2(c-3)$, or, equivalently, $c=m^2/2+3$, so $c-3+\frac{1}{2} \vec p^2=(1/2) (m^2+\vec p^2)$.

We introduce a ``cutoff" $\Lambda$ and consider only $|p|\leq \Lambda$.
So, we approximate \cref{consistency} by
\be
\label{singint}
F(m^2)\equiv \frac{1}{2}\int_{|\vec p| \leq \Lambda} (m^2+ \vec p^2)^{1/2}  \,\frac{{\rm d}^2 p}{(2\pi)^2}=\frac{h}{b},
\ee
where $F(m^2)$ is defined to be the given integral.

The integral $F(m^2)$ has some dependence on cutoff $\Lambda$ (it is ``ultraviolet divergent"), so it is convenient to remove this dependence by differentiating twice with respect to $m^2$.  We get
$$\frac{\partial^2 F(m^2)}{\partial (m^2)^2}=-\frac{1}{8} \int_{|\vec p|\leq \Lambda} (m^2+\vec p^2)^{3/2} \,\frac{{\rm d}^2 p}{(2\pi)^2}.$$
This integral now diverges at small $|\vec p|$ for $m=0$.  This divergent is cutoff for nonzero $m$ and the integral is then (for $m<<\Lambda^2$) approximately equal to $-C_1/m=-C_1/(m^2)^{1/2}$ for some $C_1>0$.
Hence, the integral in \cref{singint} behaves for small $m$ like $C+C' m^2-C''m^3+\ldots$ for some constants $C,C',C''>0$.

So, keeping only the leading terms for small $m$, we have that $C+C'm^2-C''m^3=h/b$ and we wish to minimize $b^2+c$, with
$c=3+m^2/2$.  So, $$b^2+c
=h^2/(C+C'm^2-C''m^3)^2+3+m^2/2.$$
This gives some function of $m$, which can be expanded for small $m$ as $c_0+c_2 m^2 + c_3 m^3+c_4 m^4+\ldots$, for some constant $c_0,c_2,c_3,c_4$.  The quantity $c_2$ depends on $h$.  Indeed, there is some critical value of $h$, $h_{\rm cr}$, above which the system is in the paramagnetic phase.  This $h_{\rm cr}$ is the value of $h$ at which $c_2$ vanishes.
For $h>h_{\rm cr}$, the quantity $c_2$ is negative.
Since $c_2$ depends linearly on $h-h_{\rm cr}$ for small $h-h_{\rm cr}$,
we are minimizing a function of $m$ of the form $c_0+d(h_{\rm cr}-h) m^2+c_3 m^3+\ldots$, for some constant $d$.

Hence, for small $h-h_{\rm cr}$, the minimum is at $m \sim h-h_{\rm cr}$, i.e., 
$m^2 \sim (h-h_{\rm cr})^2$.
This in fact is the scaling dependence of mass on distance from critical point which occurs for the $O(N)$ vector model in the limit of large $N$ in $2+1$ dimensions.

We leave it to the reader to verify that the singular behavior of the ground state energy matches that from the large-$N$ vector model.

\section{Beyond the Hamiltonian Formulation}
\label{secnonlocal}
The sum-of-squares method is tied to the Hamiltonian formulation of quantum mechanics.
One might wonder: suppose one has a relativistically invariant field theory; is there some covariant version of the sum-of-squares method?  One could of course build transfer matrices perpendicular to any spacelike surface and apply the sum-of-squares method to that transfer matrix.

However, one might wonder, is there some variant of the sum-of-squares method which is more closely related to the path integral, or action, formulation of quantum mechanics?
Since the action formalism is often applied to systems where there is some nonlocal action in time after integrating out degrees of freedom, it is interesting to consider the possibility of applying a sum-of-squares to some system that is nonlocal in time.
That is what we consider in this section.

To describe such a system, we define a function $Z(\beta,g)$ analogous to the partition function in the following way
\begin{eqnarray}\label{Znonlocdef}
Z(\beta,g)={\rm tr}\Bigl( {\cal T} \exp(
-\int_0^{\beta} H(\tau)\, {\rm d}\tau
-g\sum_a \int_0^{\beta} \int_0^{\beta} \Delta_a^\dagger(\tau) \Delta_a(\tau') F_a(\tau-\tau') {\rm d}\tau \, {\rm d}\tau'
)
\Bigr).
\end{eqnarray}

This expression will require some explanation.  Here $\beta,g$ are non-negative real scalars.
We assume that some finite dimensional Hilbert space is given, and $H$ is some Hermitian operator on this Hilbert space, and the $\Delta_a$ is some (possibly non-Hermitian) operator, where $a$ is some discrete index ranging over some given finite set.  The functions
$F_a(\cdot)$ are some non-negative functions which are periodic in $\beta$ so that $F(x+\beta)=F(x)$.  The notation ${\cal T}{\rm tr}(\cdot)$ 
means a ``time-ordered trace".
Let us define this formally for those unfamiliar with this notation: first, one formally expands the exponential in a power series.  Then, any given term in the power series is some time-ordered trace of integrals
over some ``time parameters" $\tau_1,\tau_2,\ldots$.   We bring these integrals outside the ``time-ordered trace" and also bring the functions $F_a$ outside the trace, i.e., for any function $f(\cdot)$ of some number of time parameters, and any operators $O_1,O_2,\ldots$ we define
\begin{eqnarray}
&&{\cal T}{\rm tr}\Bigl(\int f(\tau_1,\tau_2,\ldots,\tau_n) O_1(\tau_1) O_2(\tau_2) \ldots O_n(\tau_n) {\rm d}\tau_1 \,{\rm d}\tau_2
\,\ldots\,{\rm d}\tau_n\Bigr)
\\ \nonumber
&\equiv &
\int f(\tau_1,\tau_2,\ldots,\tau_n) {\cal T}{\rm tr}\Bigl(O_1(\tau_1) O_2(\tau_2) \ldots O_n(\tau_n)\Bigr) {\rm d}\tau_1 \,{\rm d}\tau_2
\,\ldots\,{\rm d}\tau_n.
\end{eqnarray}
To define the time-ordered trace, we define
$${\cal T}{\rm tr}(O_1(\tau_1) O_2(\tau_2) \ldots O_n(\tau_n)\equiv
{\rm tr}(O_{\pi(1)} O_{\pi(2)} \ldots O_{\pi(n)}),$$
where $\pi$ is a permutation such that
$$\tau_{\pi(1)}<\tau_{\pi(2)}<\ldots<\tau_{\pi(n)}.$$
If some time parameters in the integral coincide (e.g. $\tau_1=\tau_2$), we leave the time-ordered trace ill-defined, but this does not contribute to the integral.
 
Thus, what we would hope is that if $g>0$ then $Z(\beta,g)\leq Z(\beta,0)$.
Indeed, suppose we take a limit in which for each $a$, the function $F_a(\tau-\tau')=\lim_{\epsilon\rightarrow 0^+}\delta(\tau-\tau'-\epsilon)$.  Here we are being slightly loose about the use of Dirac $\delta$-functions but the reader can easily replace them with some sufficiently sharply peaked functions if desired.
Then, this is the same as considering the partition function of Hamiltonian
$H+g\sum_a \Delta_a^\dagger \Delta_a$ at inverse temperature $\beta$, i.e., in this case
$Z(\beta,g)={\rm tr}(\exp(-\beta (H+g\sum_a \Delta_a^\dagger \Delta_a))$.
This may be seen to be a non-increasing function of $g$ as in this case
$\partial \ln Z(\beta,g)=-g \sum_a \langle \Delta_a^\dagger \Delta_a \rangle\leq 0$, where
$\langle \ldots \rangle$ denotes the thermal expectation value.

However, what we will argue is that if we consider more general $F(\cdot)$, then we may have $Z(\beta,g)>Z(\beta,0)$ for some small nonzero $g>0$.
Indeed, we will do this in the case that $\Delta_a=\Delta^\dagger_a$ and that the functions $F_a$ are even.
We give this and some other examples in \cref{somece}; these examples are probably well-known but I do not know a reference.
Then in \cref{sosnonlocal}, we give some sufficient conditions on the functions $F$ to have $Z(\beta,g)\leq Z(\beta,0)$.

\subsection{Some Counter-Examples}
\label{somece}
The first example has a two-dimensional Hilbert space, corresponding to a single qubit.
We let $H= V Z$ where $V$ is a scalar.  The index $a$ takes only a single possible value, so we omit the subscripts on $\Delta$ and $F$.  We let $\Delta=\Delta^\dagger=X$.  Finally, we choose $$F(x)=\sum_{n} (\delta(x-\tau_0+n\beta)+\delta(x+\tau_0+n\beta)),$$ for some $\tau_0>0$, where the sum is over integer $n$ to make the function periodic in $\beta$ as required.

Then, let us take some fixed $V>0$, take $|g|<<1$ with $g>0$, and take $V^{-1}<<\tau_0<<\beta$.  To analyze this, note that if $g=0$, then it simply describes a Hamiltonian $Z$ , whose ground state is the spin down state.  Suppose we expand $Z(\beta,g)$ in powers of $g$.  The first order term in $g$ is proportional to $\exp(-V\tau_0)$ as each operator $\Delta_a$ flips the spin from ground state to excited state, and the spin does not flip back until time $\tau_0$ later.  However, at second order, we can obtain contributions which are not exponentially suppressed in $\exp(-V\tau_0)$, but only suppressed by a power of $V$.  Indeed, consider a term 
$$\Delta(\tau_1) \Delta(\tau_1') F(\tau_1-\tau_1')
\Delta(\tau_2) \Delta(\tau_2') F(\tau_2-\tau_2').$$
Then, it may be that $V |\tau_1'-\tau_2'|<<1$ and $V|\tau_1-\tau_2|<<1$.  For example, we might have
$\tau_1=0, \tau_2=\epsilon, \tau'_1=\tau_0,\tau'_2=\tau_0+\epsilon$, for some small $\epsilon$.
In this case, the suppression is only exponentially small in $V\epsilon$, and integrating over
$\epsilon$ simply gives a factor $V^{-1}$ so the overall contribution is of order $\beta g^2/V$.

We may then choose the parameters $g,V,\tau_0$ so that this particular second order contribution is exponentially larger than the first order contribution, and yet still have $g^2/V \ll 1$.  
In this case, we expect that the partition function is growing exponentially in $\beta g^2/V$, i.e., we may have $Z(\beta,g)>Z(\beta,0)$.

We omit any formal proof that the partition function is growing exponentially in $\beta g^2/V$ (though it is probably not difficult), but instead give the standard argument.
First, 
of course, there is another second order contribution to $Z(\beta,g)$ which is proportional $\beta^2$.  Indeed, this contribution is $1/2$ times the square of the first order contribution.  However, the standard way to deal with this is to perform a series expansion for $\ln(Z(\beta,g))$, and then the only second order contribution is the positive contribution proportional to $\beta g^2/V$ and we may choose parameters so that the sum of the first two contributions is positive.

In this example, the sign of the second order contribution is positive and larger than the first order contribution.  It is easy to see that the sign of the first order contribution is always negative.
It is interesting to note though that we may also have a negative sign for the second order contribution to $\log(Z)$.
To do this, we take a four-dimensional Hilbert space, corresponding to two qubits.
We let $H= V Z_1$ where $V$ is a scalar.  We let $\Delta_1=X_1 X_2$, let $\Delta_2=X_1 Y_2$, and let $\Delta_3=X_1 Z_2$.  Finally, we choose $F_a(x)=F(x)$, where $F(x)$ is as before.
Then, the first order contribution is exponentially suppressed as before, but one may verify that the sign of the second order contribution is negative.

\subsection{Sufficient Conditions}
\label{sosnonlocal}
Having seen that it is not enough to have $F_a(\tau-\tau')$ be a non-negative function to have $Z(\beta,g)\leq Z(\beta,0)$, we now give some sufficient conditions.  Rather than just stating the conditions and then proving that they are sufficient, we instead derive them in some sense.

Consider the following alternative definition of a function $Z(\beta,g)$.  The Hilbert space is a tensor product of a finite dimensional Hilbert space on which $H,\Delta_a$ act and also some additional harmonic oscillators, one such oscillator for each choice of the index $a$.
We let
\be
\label{Zbgaux}
Z(\beta,g)={\cal T} {\rm tr}\Bigl(\exp[-\int_0^\beta H(\tau)+H_{\rm harmonic}+i\sqrt{g} V(\tau) \, {\rm d}\tau]\Bigr),
\ee
where the trace is over both finite dimensional and harmonic oscillator Hilbert space,
where
$$
H_{\rm harmonic}=\sum_a \epsilon_a b^\dagger_a b_a,
$$
where
where $b_a,b_a^\dagger$ are creation and annihilation operators on the given harmonic oscillator and
$\epsilon_a>0$ are real scalars, and where
$$V(\tau)\equiv \sum_a \exp(i \omega_a \tau) \Delta_a(\tau) b^\dagger_a(\tau) + \exp(-i \omega_a \tau) \Delta_a^\dagger(\tau) b_a(\tau)),$$
where $\omega_a$ are some real numbers which are integer multiple of $2\pi/\beta$, and $\Delta_a$ are some operators.

Note that if all $\omega_a$ are equal to $0$, then we do not need to use time-ordered traces.  In this case we would simply have $Z(\beta,g)={\rm tr}(\exp[-\beta(H+\sum_a b_a^\dagger b_a+i\sqrt{g}(\sum_a F_a(\tau) b_a^\dagger+F_a^\dagger b_a))])$.

For $g\geq 0$, the term $i\sqrt{g}$ is anti-Hermitian, regardless of $\omega_a$.
We claim then that
$Z(\beta,g)\leq Z(\beta,0)$.  Indeed, this follows because we can (to any desired accuracy) approximate
$Z(\beta,g)$ by a Trotter-Suzuki decomposition by
$$Z(\beta,g) \approx {\rm tr}\Bigl(\exp(-\frac{\beta}{n}(H+H_{\rm harmonic}) U_1 
\exp(-\frac{\beta}{n}(H+H_{\rm harmonic}) U_2 \ldots \exp(-\frac{\beta}{n}(H+H_{\rm harmonic}) U_n
\Bigr),$$
where each unitary matrix $U_j=\exp[i\sqrt{g}(\beta/n) 
\sum_a \exp(-i \omega_a \tau) \Delta_a b^\dagger_a + \exp(+i \omega_a \tau) \Delta_a^\dagger b_a)],$
and
where the error in the Trotter-Suzuki approximation tends to zero as the integer $n$ tends to infinity.
Then, by a generalization of von Neumann's trace inequality due to \cite{fan1951maximum}, this is bounded by
$Z(\beta,0)$; see also theorem 20B.2 of \cite{marshall1979inequalities} for this generalization.  The generalization is that given any trace ${\rm tr}(A_1 U_1 A_2 U_2 \ldots A_m U_m)$, where $U_m$ are unitary matrices, and the $A_j$ have singular values $\sigma_1(A_j)\geq \sigma_2(A_j)\geq  \ldots \geq 0$, the trace is bounded by $\sum_{i} \sigma_i(A_1) \sigma_i(A_2) \ldots \sigma_i(A_m)$.

However, we may then, using standard field theory techniques, integrate out the harmonic oscillators, leaving an expression that involves only the finite dimensional Hilbert space.  The result is
$$
Z(\beta,g)=Z_{\rm hamornic}(\beta) \times {\rm tr}_{\rm qudit}\Bigl( {\cal T} \exp(
\int_0^{\beta} H(\tau)\, {\rm d}\tau
+g\sum_a \int_0^{\beta} \int_0^{\beta} \Delta_a^\dagger(\tau) \Delta_a(\tau') \exp(i \omega(\tau-\tau')) G_a(\tau-\tau') {\rm d}\tau \, {\rm d}\tau'
)
\Bigr),
$$
where the notation ${\cal T}{\rm tr}_{\rm qudit}(\cdot)$ indicates a time-ordered trace just over the finite dimensional Hilbert space,
where $Z_{\rm harmonic}(\beta)={\rm tr}_{\rm harmonic}(\exp(-\sum_a \epsilon_a b^\dagger_a b_a))$ is the partition function of the harmonic oscillators (here the notation indicates that the trace is just over the harmonic oscillator Hilbert space),
and where
\begin{eqnarray}
G_a(\tau-\tau')&=&Z_{\rm harmonic}(\beta)^{-1} 
{\rm tr}_{\rm harmonic}\Bigl(\exp(-\sum_a \epsilon_a b^\dagger_a b_a)\, b_a(\tau) b^\dagger_a(\tau')\Bigr)
\\ \nonumber
&\equiv &
\langle b_a(\tau) b^\dagger_a(\tau')\rangle_\beta,
\end{eqnarray}
where the notation $\langle \ldots \rangle_\beta$ denotes a thermal expectation value at inverse temperature $\beta$ as defined by the above equation.  We use the symbol $G_a$ to denote this function because it is a Green's function of the harmonic oscillator.

For finite $\beta$, the Green's function is some periodic function of $\beta$.  However, it is interesting to consider the limit of large $\beta$.  In this case, $G_a$ converges to $\theta(\tau-\tau')\exp(-\epsilon_a (\tau-\tau'))$, where $\theta(\cdot)$ is a step function.  That is, when $\tau-\tau'$ is small compared to $\beta$, it converges to a step function.  We remind the reader that we are being slightly careless about $\delta$-functions and step functions, but it is not difficult to fill in the details.

Thus, identifying $F_a(\tau-\tau')=\exp(i \omega_a(\tau-\tau')) \exp(-\epsilon_a(\tau-\tau')) \theta(\tau-\tau')$ gives us 
a choice of functions $F_a$ for which $Z(\beta,g)$ as in \cref{Znonlocdef} has the property $Z(\beta,g)\leq Z(\beta,0)$ for $g>0$.

Of course, we can also use this trick of integrating out harmonic oscillators to obtain functions
$F_a(\tau-\tau')$ which are equal to $\exp(-\epsilon_a(\tau-\tau')) \theta(\tau-\tau')\sum_b W_b \exp(i \omega_{a,b} (\tau-\tau'))$ where the sum is over some discrete index $b$, where $W_b>0$, and where $\omega_{a,b}$ is some function of $a$ and $b$, by indexing the harmonic oscillators with a pair of indices $a,b$.

By doing this and taking a large number of terms in the sum while taking $\epsilon_a \rightarrow 0^+$,
we 
expect that we will have the property
$Z(\beta,g)\leq Z(\beta,0)$ for $g>0$ in the limit of large $\beta$
whenever
$F_a(\tau-\tau')=\theta(\tau-\tau') f_a(\tau-\tau')$ for any choices of functions $f_a$
which have Fourier transforms which are non-negative and sufficiently well-behaved (we leave details of what this would mean to the reader!).

\section{On Classical Methods for SYK Ground States}
\label{sykground}
In this section, we consider two classical variational methods for approximating the ground state energy of the SYK model.  One method is the Lanczos algorithm, starting with a Gaussian wavefunctions (we explain Gaussian states and wavefunctions in more detail below), and the other method is a sum of Gaussian wavefunctions.  We prove limitations on the power of these methods.

\subsection{Background}
The SYK model\cite{SY93,Kit15} is a model of fermions with randomly chosen interactions.
See \cite{rosenhaus2019introduction,GV16,GJV18,FTW19,FTW18,FTW20}.

In this paper, we largely consider the degree-$4$ SYK model.  In this case, the Hamiltonian is
$$H=\sum_{i,j,k,l} J_{ijkl} \gamma_i \gamma_j \gamma_k \gamma_l,$$
where $\gamma_i$ are Majorana operators obeying the anti-commutation relations $\{\gamma_i,\gamma_j\}=2\delta_{i,j}$, with $i\in \{1,2,\ldots,n\}$, with $n$ even, and where
the entries of the tensor $J$ are independent Gaussians, up to the requirement that $J$ be totally anti-symmetric in its indices.

More generally, one can consider a degree-$4$ SYK model for even $q>4$.  In this case, we mean a sum of degree-$q$ monomials in Majorana variables, with Gaussian random coefficients, with variance chosen so that the expected sum-of-squares of coefficients is equal to $1$.  We discuss this briefly later, but if not otherwise specified, we mean the degree-$4$ model.

We choose the variance of the Gaussians so that the $\ell_2$ norm of $J$ (i.e., the square-root of sum-of-squares of its entries) is a constant, independent of $n$.  This is a different normalization that considered in physics, where instead for the degree-$4$ model the $\ell_2$ norm is of order $\sqrt{n}$, but is convenient for us and was used in Ref.~\cite{hastings2022optimizing}.

Also, we will (following \cite{hastings2022optimizing}) consider approximating the state with most positive eigenvalue (i.e., the highest ezcited state), rather than the state with most negative eigenvalue (i.e., the ground state).  The distribution of $J$ is invariant under change of sign, so this has no effect, but it avoids some signs later.

Mathematical physics results predict that with this normalization, the largest eigenvalue is proportional to $\sqrt{n}$ with high probability, and even predict the leading coefficient, though it is not proven.
In \cite{FTW19}, it is proven that with high probability the largest eigenvalue is $\cO(\sqrt{n})$ with high probability.  In \cite{hastings2022optimizing}, it is also proven that with high probability, the largest eigenvalue is $\Omega(\sqrt{n})$, thus proving that the eigenvalue is $\theta(\sqrt{n})$ with high probability.

The eigenstate with largest eigenvalue of the SYK model is predicted to be highly entangled\cite{liu2018quantum}.  In this regard, it is interesting to see to what extent one can find variational states which still have a large expectation value for the SYK Hamiltonian.  In \cite{hastings2022optimizing}, it was shown that, with high probablity, one could efficiently \emph{on a quantum computer} construct a quantum variational state which had energy which is $\theta(\sqrt{n})$, where by ``energy" of a state, we simply mean the expectation value of the SYK Hamiltonian.

However, suppose we restrict to variational states whose energy can be efficiently evaluated on a \emph{classical computer}.
In this case, 
\cite{haldar2021variational} proved an important negative result: with high probability, for any Gaussian state, the expectation value of the SYK Hamiltonian is $\cO(1)$, which has a different scaling with $n$ than the largest eigenvalue.

In this section, we prove further results.  
Our main result is to bound the expectation value of the energy of a \emph{sum} of Gaussian wavefunctions,  i.e., some sum
$\sum_i a_i \psi_i$, where $\psi_i$ are Gaussian wavefunctions, under
an assumption explained later on the norm.  
Note, a sum of polynomially many Gaussian wavefunctions is
an important  class of states where one can efficiently evaluate the energy on a classical computer is a sum of Gaussian wavefunctions\cite{bravyi2017complexity,boutin2021quantum}.  Importantly, these Gaussian wavefunctions do \emph{not} need to be orthogonal to each other.  If they are orthogonal, our norm assumption is fulfilled, so long as the total number of Gaussians is sufficiently small; the number we allow is exponentially large in a power of $n$.  Even if the wavefunctions are not orthogonal, the norm assumption may be fulfilled.
As a corollary, we prove
a limitation on the power of Lanczos methods starting with a Gaussian wavefunction.
The Lanczos method is variational method within the subspace, called a ``Krylov space", which is the
span of $\psi,H\psi,H^2\psi,\ldots,H^k\psi$, for some finite $k$.  When we say we ``start" with a Gaussian wavefunction, we mean that $\psi$ is a Gaussian wavefunction (explained next).

Before doing this,
let us define what we mean by Gaussian states and give some mathematical background on Wick's theorem.
Here a ``state" refers to a density matrix.  Gaussian states are those states in which expectation values of
any product of Majorana operators are determined by Wick's theorem (below).  The pure Gaussian states are precisely the states which are ground states of Hamiltonians which have unique ground states and which are quadratic in Majorana operators.
See Ref.~\cite{bravyi2004lagrangian} for more details.

We use the term ``wavefunction" to mean a vector $\psi$ in the Hilbert space describing the given quantum system, so that for a normalized wavefunction $|\psi\rangle$, the projector $|\psi\rangle \langle \psi|$ is a state.  We say that $\psi$ is a Gaussian wavefunction if the corresponding projector is a Gaussian state.

Let us briefly review Wick's theorem.
Let $M$ be a matrix with matrix elements $M_{lm}=\langle \gamma_l \gamma_m\rangle$, where $\langle \ldots \rangle$
denotes the expectation value in a given Gaussian state.  In general we have
$M=I+iB$, where $I$ is the identity matrix and $B$ is a real anti-symmetric matrix, with eigenvalues of $iB$ bounded by
$1$ in absolute value.  If the Gaussian state is pure, then $B^2=-I$, and the state is a ground state of a quadratic Majorana Hamiltonian.

Any higher order expectation value
$\langle \gamma_{i_1} \gamma_{i_2} \ldots \gamma_{i_{2m}} \rangle$ in a Gausian state can be computed as follows.
Consider all possible ways of pairing the $2m$ different Majorana operators with each other.  There are
$(4k)!!\equiv (4k-1)\cdot (4k-3)\cdot \ldots \cdot 1$ such pairings.  
We will regard a pairing of these Majorana operators as a pairing of the integers $1,2,\ldots,2m$.
Each pairing defines some
sequence of pairs $(a_1,b_1),(a_2,b_2),\ldots,(a_m,b_m)$ with $a_1<a_2<\ldots<a_m$ and $b_1<b_2<\ldots<b_m$.
For each pairing, consider the product  $$\prod_{j} M_{i_{a_j},i_{b_j}}.$$
Then, sum this product over all pairings, with a sign equal to the sign of the permutation from
the sequence $a_1,b_1,a_2,b_2,\ldots,a_m,b_m$ to the sequence $1,2,3,\ldots,2m$.

Remark: often when Wick's theorem is given for Majorana fermions, it is assumed that all $i_1,\ldots,i_{2m}$ are distinct from each other.  Of course, we can always reduce to this case by using the Majorana anti-commutation relations.  In this case, the sum over pairings has a nice representation as a Pfaffian.

\subsection{Main Results}
We now prove the main results.

The results are largely corollaries of the following:
\begin{theorem}
\label{gexpecthm}
With high probability, for $H$ drawn from the SYK distribution,
the expectation value of $H^k$, for integer $k\geq 0$, in any Gaussian state is bounded by
$$(\cO(k))^{2k}.$$
\begin{proof}
We compute expectation value of $H^k$ by Wick's theorem.  We bound the product $\prod_{j} M_{i_{a_j},i_{b_j}}$ in each pairing, and then sum over pairings using a triangle inequality.

To bound a pairing, as in \cite{hastings2022optimizing}, 
note that a pairing can be regarded as a tensor network.  There are degree $4$ vertices corresponding to the four-index tensor $J$.  There are $k$ such vertices.  We join these tensors in the way corresponding to the given pairing, and then in each edge we insert a degree-$2$ vertex, corresponding to the matrix $M$.  We will write this tensor network as a particular product of vectors and matrices. 

Let us
define $J_{2,2}^{mat}$ to be an $n^2$-by-$n^2$ matrix, with rows (and columns) indexed by pairs $(i,j)$ with $i,j\in \{1,\ldots,n\}$, with matrix element $(J_{2,2}^{mat})_{(i,j),(k,l)}=J_{i,j,k,l}$.
The matrix $J_{2,2}^{mat}$ was called simply $J^{mat}$ in \cite{hastings2022optimizing}.  We use this additional notation because we also introduce matrices $J_{p,4-p}^{mat}$ which are defined to be $n^p$-by-$n^{4-p}$ matrices, where the rows are indexed by $p$ integers from $\{1,\ldots,n\}$ and the columns are indexed by $4-p$ such integers, with
matrix elements defined in the obvious way:
$$\Bigl(J_{p,4-p}^{mat}\Bigr)_{(i_1,\ldots,i_p),(j_1,\ldots,j_{4-p})}=J_{i_1,\ldots,i_p,j_1,\ldots,j_{4-p}}.$$

Then we assign each of the degree $2$ vertices of the network a label, either ``left" or ``right".  
Then, the value of the tensor network can be written as some some expectation value as follows.  Let $M_{vec}$ be a vector in ${\mathbb C}^{n^2}$ given by regarding $M$ as a vector.  Then, the contraction of the tensor network
equals
$$\langle \pi_L (M_{vec})^{\otimes N_L} | (J_{0,4}^{mat})^{\otimes k_{0,4}} \otimes
(J_{1,3}^{mat})^{\otimes k_{1,3}} \otimes
(J_{2,2}^{mat})^{\otimes k_{2,2}} \otimes
(J_{3,1}^{mat})^{\otimes k_{3,1}}\otimes
(J_{4,0}^{mat})^{\otimes k_{4,0}}
 | \pi_R (M_{vec})^{\otimes N_R} \rangle.$$
Let us explain the meaning of this.  
We let $N_L$ be the number of left vertices and $N_R$ be the number of right vertices,
with $N_L+N_R=2k$.
Here we are using the bra-ket notation simply as a way of writing a vector-matrix-vector product, with the bra vector in ${\mathbb C}^{n^{2N_L}}$ and the ket vector in
 ${\mathbb C}^{n^{2N_R}}$.  Here we label basis vectors of ${\mathbb C}^{n^{2N_L}}$ by $2N_L$ indices, each in $\{1,\ldots,n\}$, and $\pi_L$ denotes an operator which applies some permutation of these indices, i.e., it maps a given basis vector to some other basis vector by permuting the indices in some given way.  We define $\pi_R$ similarly to permute indices.
The number $k_{p,4-p}$ is equal to the number of degree $4$ vertices in the tensor network for which $p$ edges connect to left vertices and $4-p$ connect to right vertices.

The advantage of this representation is that we can apply norm bounds.  
Let $v_L=\pi_L (M_{vec})^{\otimes N_L}$ and let $v_R= \pi_R (M_{vec})^{\otimes N_R}$.
We have
$|v_L|\leq \cO(n^{N_L/2})$ and $|v_R| \leq \cO(n^{N_R/2})$, where we use the $\ell_2$ norm.

Standard random matrix theory bounds show that, with high probability, $\Vert J_{2,2}^{mat} \Vert \leq \cO(1/n)$, and $\Vert J_{1,3}^{mat} \Vert = \Vert J_{3,1}^{mat}\Vert \leq \cO(1/\sqrt{n})$ and
$\Vert J_{0,4}^{mat} \Vert = \Vert J_{4,0}^{mat} \Vert \leq \cO(1)$, where $\Vert \ldots \Vert$ denotes the operator norm.
In the rest of the proof, we will assume that these bounds on operator norms hold.

So, the value of the tensor network is bounded in absolute value by
$$\cO(n)^{(N_L+N_R)/2} \cO(n^{-1})^{k_{2,2}} \cO(n^{-1/2})^{k_{1,3}+k_{3,1}} \cO(1)^{k_{0,4}+k_{4,0}}=
\cO(n)^{k-k_{2,2}-k_{1,3}/2-k_{3,1}/2}
\cO(1)^{k}.$$

This holds for any labelling.  We now choose a labelling.
We claim that it is possible to label the degree-$2$ vertices so that each each degree-$4$ vertex has two left neighbors and two right neighbors, i.e., so that $k_{2,2}=k$.  Then, the value of the tensor network is bounded in
absolute value by
$$\cO(1)^{k}.$$
Summing over all pairings, and using $4k!!\leq (4k)^{2k}$, the theorem follows.

To prove the claim on the labelling of vertices,
it is convenient to consider the multigraph obtained by connecting the degree-$4$ vertices according to the given pairing, without inserting the degree-$2$ vertices in each edge.  In this case, the question is whether we can color the edges of a regular (i.e., all vertices have the same degree) degree-$4$ multigraph so that each vertex has two edges of each color attached to it. 
This is possible as follows\footnote{I thank R. O'Donnell for this proof of the existence of such a coloring.}:
since the multigraph has even degree, each connected component is Eulerian.  Take an Eulerian cycle in each connected component and alternately color the edges.  Since the degree of each vertex is zero mod $4$, the total number of edges is even in each connected component, and so Eulerian cycle has even length, making this alternation of coloring possible.
\end{proof}
\end{theorem}

Remark: one may also consider the case of the degree-$q$ SYK model for even $q>4$.  There are some modifications needed to the above proof.  We may define an analogous $J^{mat}_{p,q-p}$.  Then, we have, with high probability that $\Vert J^{mat}_{q/2,q/2} \Vert \leq \cO(n^{-q/4})$.
Also, if $q=0 \mod 4$, then the analogous coloring argument works, and it is possible to color edges of a regular degree-$q$ multigraph so that each vertex has $q/2$ edges of each color attached to it.  However, if $q=2 \mod 4$, then the coloring argument need not work since the length of the Eulerian cycle is odd if there are an odd number of vertices in a given connected component.  In that case, if the number of vertices in a component is odd,
it is possible to color so that all but one vertex in each connected component has $q/2$ edges of each color attached to it, and the remaining vertex may be colored so that it has $q/2+1$ edges of one color and $q/2-1$ edges of the other color.  Indeed, this may be the best possible; for example, consider the case $q=6$ and consider a multigraph with three vertices of degree $6$, with $3$ edges connecting each pair of vertices.

Hence,
\begin{corollary}
\label{projcorr}
Let $H$ be drawn from the SYK distribution.  Let $c$ be any positive constant.
Then, with high probability, for any Gaussian state $\rho$ the projection of $\rho$ onto the
eigenspace of $H$ with eigenvalue $\geq c\sqrt{n}$
is exponentially small in $n^{1/4}$, i.e., the projection is bounded by $(c')^{n^{1/4}}$ for some $c'<1$.
\begin{proof}
Let $\Pi$ project onto the eigenspace of $H$ with eigenvalue $\geq c \sqrt{n}$
Pick $k$ even.  Then ${\rm tr}(\rho H^k) \geq {\rm tr}(\rho \Pi) c^k n^{k/2}$.  By \cref{gexpecthm}, with high
probability we have
${\rm tr}(\rho H^k)\leq (\cO(k))^{2k}$.  So
${\rm tr}(\rho \Pi) \leq  (\cO(k))^{2k} c^{-k} n^{-k/2}=(\cO(k) n^{-1/4} c^{-1/2})^{2k}$.
We pick $k$ to be the largest even integer less than
$c'' n^{1/4}$ for some $c''>0$.  Picking $c''$ small enough, the result follows.
\end{proof}
\end{corollary}

So,
\begin{corollary}
\label{sumgaussassumpt}
Let $H$ be drawn from the SYK distribution.
Given any sum of polynomially many Gaussian wavefunctions, of the form
$\Psi\equiv\sum_i a_i \psi_i$, with $\psi_i$ being Gaussian wavefunctions, then, with high probability,
if $\log(\sum_i |a_i|/|\Psi|)$ is $o(n^{1/4})$,
then $\langle \Psi | H |\Psi \rangle/|\Psi|^2=o(\sqrt{n})$.
Remark: colloquially one may say that the condition is that $\sum_i |a_i|/|\Psi|$ is not exponentially large in $n^{1/4}$.

Further, if the Gaussians are orthogonal to each other, and if the number of Gaussian wavefunctions in the sum
is some $k$ with $\log(k)=o(n^{1/4})$, then $\langle \Psi | H |\Psi \rangle/|\Psi|^2=o(\sqrt{n})$.
\begin{proof}
For any $c>0$,
the norm of the projection of $\psi_i$ onto the eigenspace with eigenvalue $\geq c\sqrt{n}$ is exponentially small in $n^{1/4}$.  Hence, by a triangle inequality, the norm of the projection of $\Psi$ onto the given eigenspace is bounded by $\sum_i |a_i|$ times something exponentially small in $n^{1/4}$.

To show the second claim, if the Gaussian wavefunctions are orthogonal, and there are $k$ wavefunctions in the sum, then $|\Psi|/\sum_i |a_i|\geq 1/\sqrt{k}.$
\end{proof}
\end{corollary}

The above corollary needs this assumption on the norm $|\Psi|/\sum_i |a_i|$ or on orthogonality of the wavefunctions.
We conjecture that this assumption is not necessary.
\begin{conjecture}
Let $H$ be drawn from the SYK distribution.
Given any sum of polynomially many Gaussian wavefunctions, the expectation value of the $H$
in the resulting state is $o(\sqrt{n})$.
\end{conjecture}

Before proving this, we give the immediate corollary:
\begin{corollary}
With high probablity, 
using
$o(n^{1/4})/\log(n)$ steps of the Lanczos algorithm or the power method, starting from a Gaussian wavefunction, produces a state whose for the SYK Hamiltonian  is
$o(\sqrt{n})$.
\begin{proof}
A Gaussian wavefunction is the ground state of a quadratic Majorana Hamiltonian which naturally defines an orthonormal basis of states.
Given such a quadratic Hamiltonian $H_{\rm quad}$, one can
pick a new basis of Majorana operators that we write as $\tg_1,\ldots,\tg_n$ such that in this
basis $$H_{\rm quad}=\sum_{j=1}^{n/2} \Bigl( 1+i\tg_{2j-1} \tg_{2j} \Bigr).$$
These operators also obey the canonical anti-comutation relations:
$$\{\tg_j,\tg_k\}=2\delta_{j,k}.$$
Then, this defines a natural orthonormal basis of states, where each such state is an
eigenstate of all the operators $i\tg_{2j-1} \tg_{2j}$ and hence is a Gaussian wavefunction.  Each such operator has eigenvalues $\pm 1$ and there are $2^{n/2}$ such states.

We say a state has $m$ excitations if there are $m$ such operators with eigenvalue $+1$ and the others all have eigenvalue $-1$.
Starting with a Gaussian wavefunction, and applying $k$ steps of the Lanczos algorithm, one can describe the resulting state as a sum of states with up to $4k$ ``excitations".
For $k=o(n^{1/4})/\log(n)$, there are $2^{o(n^{1/4})}$ such states, and so the result follows from
\cref{sumgaussassumpt}.
\end{proof}
\end{corollary}

Note, for only logarithmically many steps of the Lanczos algorithm, the natural basis of states used in the above proof means that one can efficiently classically apply the Lanczos algorithm.  However, we do not know any way to efficiently compute a super-logarithmic number of Lanczos steps.

\bibliography{qsos-ref}

\begin{thebibliography}{42}
\expandafter\ifx\csname natexlab\endcsname\relax\def\natexlab#1{#1}\fi
\expandafter\ifx\csname bibnamefont\endcsname\relax
  \def\bibnamefont#1{#1}\fi
\expandafter\ifx\csname bibfnamefont\endcsname\relax
  \def\bibfnamefont#1{#1}\fi
\expandafter\ifx\csname citenamefont\endcsname\relax
  \def\citenamefont#1{#1}\fi
\expandafter\ifx\csname url\endcsname\relax
  \def\url#1{\texttt{#1}}\fi
\expandafter\ifx\csname urlprefix\endcsname\relax\def\urlprefix{URL }\fi
\providecommand{\bibinfo}[2]{#2}
\providecommand{\eprint}[2][]{\url{#2}}

\bibitem[{\citenamefont{Hastings and O'Donnell}(2022)}]{hastings2022optimizing}
\bibinfo{author}{\bibfnamefont{M.~B.} \bibnamefont{Hastings}} \bibnamefont{and}
  \bibinfo{author}{\bibfnamefont{R.}~\bibnamefont{O'Donnell}}, in
  \emph{\bibinfo{booktitle}{Proceedings of the 54th Annual ACM SIGACT Symposium
  on Theory of Computing}} (\bibinfo{year}{2022}), pp.
  \bibinfo{pages}{776--789}.

\bibitem[{\citenamefont{Hastings}(2022)}]{hastings2022perturbation}
\bibinfo{author}{\bibfnamefont{M.~B.} \bibnamefont{Hastings}},
  \bibinfo{journal}{arXiv preprint arXiv:2205.12325}  (\bibinfo{year}{2022}).

\bibitem[{\citenamefont{Bonner and Fisher}(1964)}]{bonner1964linear}
\bibinfo{author}{\bibfnamefont{J.~C.} \bibnamefont{Bonner}} \bibnamefont{and}
  \bibinfo{author}{\bibfnamefont{M.~E.} \bibnamefont{Fisher}},
  \bibinfo{journal}{Physical Review} \textbf{\bibinfo{volume}{135}},
  \bibinfo{pages}{A640} (\bibinfo{year}{1964}).

\bibitem[{\citenamefont{Schollw{\"o}ck}(2011)}]{schollwock2011density}
\bibinfo{author}{\bibfnamefont{U.}~\bibnamefont{Schollw{\"o}ck}},
  \bibinfo{journal}{Annals of physics} \textbf{\bibinfo{volume}{326}},
  \bibinfo{pages}{96} (\bibinfo{year}{2011}).

\bibitem[{\citenamefont{O'Donnell}(2017)}]{OD17}
\bibinfo{author}{\bibfnamefont{R.}~\bibnamefont{O'Donnell}}, in
  \emph{\bibinfo{booktitle}{Proceedings of the 8th annual Innovations in
  Theoretical Computer Science Conference (ITCS)}} (\bibinfo{year}{2017}).

\bibitem[{\citenamefont{Goemans and Williamson}(1995)}]{GW95}
\bibinfo{author}{\bibfnamefont{M.}~\bibnamefont{Goemans}} \bibnamefont{and}
  \bibinfo{author}{\bibfnamefont{D.}~\bibnamefont{Williamson}},
  \bibinfo{journal}{Journal of the ACM} \textbf{\bibinfo{volume}{42}},
  \bibinfo{pages}{1115} (\bibinfo{year}{1995}).

\bibitem[{\citenamefont{Charikar and Wirth}(2004)}]{CW04}
\bibinfo{author}{\bibfnamefont{M.}~\bibnamefont{Charikar}} \bibnamefont{and}
  \bibinfo{author}{\bibfnamefont{A.}~\bibnamefont{Wirth}}, in
  \emph{\bibinfo{booktitle}{Proceedings of the 45th annual Symposium on
  Foundations of Computer Science (FOCS)}} (\bibinfo{organization}{IEEE},
  \bibinfo{year}{2004}), pp. \bibinfo{pages}{54--60}.

\bibitem[{\citenamefont{Coleman}(1963)}]{coleman1963structure}
\bibinfo{author}{\bibfnamefont{A.~J.} \bibnamefont{Coleman}},
  \bibinfo{journal}{Reviews of Modern Physics} \textbf{\bibinfo{volume}{35}},
  \bibinfo{pages}{668} (\bibinfo{year}{1963}).

\bibitem[{\citenamefont{Erdahl}(1978)}]{erdahl1978representability}
\bibinfo{author}{\bibfnamefont{R.~M.} \bibnamefont{Erdahl}},
  \bibinfo{journal}{International Journal of Quantum Chemistry}
  \textbf{\bibinfo{volume}{13}}, \bibinfo{pages}{697} (\bibinfo{year}{1978}),
  \urlprefix\url{https://doi.org/10.1002%2Fqua.560130603}.

\bibitem[{\citenamefont{Percus}(1978)}]{percus1978role}
\bibinfo{author}{\bibfnamefont{J.}~\bibnamefont{Percus}},
  \bibinfo{journal}{International Journal of Quantum Chemistry}
  \textbf{\bibinfo{volume}{13}}, \bibinfo{pages}{89} (\bibinfo{year}{1978}).

\bibitem[{\citenamefont{Mazziotti and Erdahl}(2001)}]{mazziotti2001uncertainty}
\bibinfo{author}{\bibfnamefont{D.}~\bibnamefont{Mazziotti}} \bibnamefont{and}
  \bibinfo{author}{\bibfnamefont{R.}~\bibnamefont{Erdahl}},
  \bibinfo{journal}{Physical Review A} \textbf{\bibinfo{volume}{63}},
  \bibinfo{pages}{042113} (\bibinfo{year}{2001}).

\bibitem[{\citenamefont{Nakata et~al.}(2001)\citenamefont{Nakata, Nakatsuji,
  Ehara, Fukuda, Nakata, and Fujisawa}}]{Nakata_2001}
\bibinfo{author}{\bibfnamefont{M.}~\bibnamefont{Nakata}},
  \bibinfo{author}{\bibfnamefont{H.}~\bibnamefont{Nakatsuji}},
  \bibinfo{author}{\bibfnamefont{M.}~\bibnamefont{Ehara}},
  \bibinfo{author}{\bibfnamefont{M.}~\bibnamefont{Fukuda}},
  \bibinfo{author}{\bibfnamefont{K.}~\bibnamefont{Nakata}}, \bibnamefont{and}
  \bibinfo{author}{\bibfnamefont{K.}~\bibnamefont{Fujisawa}},
  \bibinfo{journal}{The Journal of Chemical Physics}
  \textbf{\bibinfo{volume}{114}}, \bibinfo{pages}{8282} (\bibinfo{year}{2001}),
  \urlprefix\url{https://doi.org/10.1063%2F1.1360199}.

\bibitem[{\citenamefont{Mazziotti}(2012)}]{Maz12}
\bibinfo{author}{\bibfnamefont{D.}~\bibnamefont{Mazziotti}},
  \bibinfo{journal}{Physical Review Letters} \textbf{\bibinfo{volume}{108}},
  \bibinfo{pages}{263002} (\bibinfo{year}{2012}).

\bibitem[{\citenamefont{Klyachko}(2006)}]{klyachko2006quantum}
\bibinfo{author}{\bibfnamefont{A.}~\bibnamefont{Klyachko}},
  \textbf{\bibinfo{volume}{36}}, \bibinfo{pages}{72} (\bibinfo{year}{2006}).

\bibitem[{\citenamefont{Helton and McCullough}(2004)}]{HM04}
\bibinfo{author}{\bibfnamefont{J.~W.} \bibnamefont{Helton}} \bibnamefont{and}
  \bibinfo{author}{\bibfnamefont{S.}~\bibnamefont{McCullough}},
  \bibinfo{journal}{Transactions of the American Mathematical Society}
  \textbf{\bibinfo{volume}{356}}, \bibinfo{pages}{3721} (\bibinfo{year}{2004}),
  ISSN \bibinfo{issn}{0002-9947},
  \urlprefix\url{https://doi.org/10.1090/S0002-9947-04-03433-6}.

\bibitem[{\citenamefont{Navascu{\'e}s et~al.}(2008)\citenamefont{Navascu{\'e}s,
  Pironio, and Ac{\'\i}n}}]{NPA08}
\bibinfo{author}{\bibfnamefont{M.}~\bibnamefont{Navascu{\'e}s}},
  \bibinfo{author}{\bibfnamefont{S.}~\bibnamefont{Pironio}}, \bibnamefont{and}
  \bibinfo{author}{\bibfnamefont{A.}~\bibnamefont{Ac{\'\i}n}},
  \bibinfo{journal}{New Journal of Physics} \textbf{\bibinfo{volume}{10}},
  \bibinfo{pages}{073013} (\bibinfo{year}{2008}).

\bibitem[{\citenamefont{Doherty et~al.}(2008)\citenamefont{Doherty, Liang,
  Toner, and Wehner}}]{DLTW08}
\bibinfo{author}{\bibfnamefont{A.}~\bibnamefont{Doherty}},
  \bibinfo{author}{\bibfnamefont{Y.-C.} \bibnamefont{Liang}},
  \bibinfo{author}{\bibfnamefont{B.}~\bibnamefont{Toner}}, \bibnamefont{and}
  \bibinfo{author}{\bibfnamefont{S.}~\bibnamefont{Wehner}}, in
  \emph{\bibinfo{booktitle}{Proceedings of the 23rd Annual IEEE Conference on
  Computational Complexity (CCC)}} (\bibinfo{year}{2008}), pp.
  \bibinfo{pages}{199--210}.

\bibitem[{\citenamefont{Pironio et~al.}(2010)\citenamefont{Pironio,
  Navascu\'{e}s, and Ac\'{\i}n}}]{PNA10}
\bibinfo{author}{\bibfnamefont{S.}~\bibnamefont{Pironio}},
  \bibinfo{author}{\bibfnamefont{M.}~\bibnamefont{Navascu\'{e}s}},
  \bibnamefont{and}
  \bibinfo{author}{\bibfnamefont{A.}~\bibnamefont{Ac\'{\i}n}},
  \bibinfo{journal}{SIAM Journal on Optimization}
  \textbf{\bibinfo{volume}{20}}, \bibinfo{pages}{2157} (\bibinfo{year}{2010}),
  ISSN \bibinfo{issn}{1052-6234},
  \urlprefix\url{https://doi.org/10.1137/090760155}.

\bibitem[{\citenamefont{Sachdev and Ye}(1993)}]{SY93}
\bibinfo{author}{\bibfnamefont{S.}~\bibnamefont{Sachdev}} \bibnamefont{and}
  \bibinfo{author}{\bibfnamefont{J.}~\bibnamefont{Ye}},
  \bibinfo{journal}{Physical Review Letters} \textbf{\bibinfo{volume}{70}},
  \bibinfo{pages}{3339} (\bibinfo{year}{1993}).

\bibitem[{\citenamefont{Kitaev}(2015)}]{Kit15}
\bibinfo{author}{\bibfnamefont{A.}~\bibnamefont{Kitaev}}, in
  \emph{\bibinfo{booktitle}{KITP Strings Seminar and Entanglement}}
  (\bibinfo{year}{2015}), vol.~\bibinfo{volume}{12}, p.~\bibinfo{pages}{26},
  \bibinfo{note}{\url{https://online.kitp.ucsb.edu/online/entangled15/kitaev/}}.

\bibitem[{\citenamefont{Blankenbecler et~al.}(1981)\citenamefont{Blankenbecler,
  Scalapino, and Sugar}}]{blankenbecler1981monte}
\bibinfo{author}{\bibfnamefont{R.}~\bibnamefont{Blankenbecler}},
  \bibinfo{author}{\bibfnamefont{D.}~\bibnamefont{Scalapino}},
  \bibnamefont{and} \bibinfo{author}{\bibfnamefont{R.}~\bibnamefont{Sugar}},
  \bibinfo{journal}{Physical Review D} \textbf{\bibinfo{volume}{24}},
  \bibinfo{pages}{2278} (\bibinfo{year}{1981}).

\bibitem[{\citenamefont{Sugiyama and Koonin}(1986)}]{sugiyama1986auxiliary}
\bibinfo{author}{\bibfnamefont{G.}~\bibnamefont{Sugiyama}} \bibnamefont{and}
  \bibinfo{author}{\bibfnamefont{S.}~\bibnamefont{Koonin}},
  \bibinfo{journal}{Annals of Physics} \textbf{\bibinfo{volume}{168}},
  \bibinfo{pages}{1} (\bibinfo{year}{1986}).

\bibitem[{\citenamefont{Levy and Clark}(2021)}]{levy2021mitigating}
\bibinfo{author}{\bibfnamefont{R.}~\bibnamefont{Levy}} \bibnamefont{and}
  \bibinfo{author}{\bibfnamefont{B.~K.} \bibnamefont{Clark}},
  \bibinfo{journal}{Physical review letters} \textbf{\bibinfo{volume}{126}},
  \bibinfo{pages}{216401} (\bibinfo{year}{2021}).

\bibitem[{\citenamefont{Osborne}(2007)}]{osborne2007simulating}
\bibinfo{author}{\bibfnamefont{T.~J.} \bibnamefont{Osborne}},
  \bibinfo{journal}{Physical review a} \textbf{\bibinfo{volume}{75}},
  \bibinfo{pages}{032321} (\bibinfo{year}{2007}).

\bibitem[{\citenamefont{Bravyi and Hastings}(2011)}]{bravyi2011short}
\bibinfo{author}{\bibfnamefont{S.}~\bibnamefont{Bravyi}} \bibnamefont{and}
  \bibinfo{author}{\bibfnamefont{M.~B.} \bibnamefont{Hastings}},
  \bibinfo{journal}{Communications in mathematical physics}
  \textbf{\bibinfo{volume}{307}}, \bibinfo{pages}{609} (\bibinfo{year}{2011}).

\bibitem[{\citenamefont{Hastings}(2004)}]{hastings2004lieb}
\bibinfo{author}{\bibfnamefont{M.~B.} \bibnamefont{Hastings}},
  \bibinfo{journal}{Physical review b} \textbf{\bibinfo{volume}{69}},
  \bibinfo{pages}{104431} (\bibinfo{year}{2004}).

\bibitem[{\citenamefont{Bravyi et~al.}(2011)\citenamefont{Bravyi, DiVincenzo,
  and Loss}}]{bravyi2011schrieffer}
\bibinfo{author}{\bibfnamefont{S.}~\bibnamefont{Bravyi}},
  \bibinfo{author}{\bibfnamefont{D.~P.} \bibnamefont{DiVincenzo}},
  \bibnamefont{and} \bibinfo{author}{\bibfnamefont{D.}~\bibnamefont{Loss}},
  \bibinfo{journal}{Annals of physics} \textbf{\bibinfo{volume}{326}},
  \bibinfo{pages}{2793} (\bibinfo{year}{2011}).

\bibitem[{\citenamefont{Polyakov}(1987)}]{polyakov1987gauge}
\bibinfo{author}{\bibfnamefont{A.~M.} \bibnamefont{Polyakov}},
  \emph{\bibinfo{title}{Gauge fields and strings}} (\bibinfo{publisher}{Taylor
  \& Francis}, \bibinfo{year}{1987}).

\bibitem[{\citenamefont{Zinn-Justin}(1998)}]{zinn1998vector}
\bibinfo{author}{\bibfnamefont{J.}~\bibnamefont{Zinn-Justin}},
  \bibinfo{journal}{arXiv preprint hep-th/9810198}  (\bibinfo{year}{1998}).

\bibitem[{\citenamefont{Fan}(1951)}]{fan1951maximum}
\bibinfo{author}{\bibfnamefont{K.}~\bibnamefont{Fan}},
  \bibinfo{journal}{Proceedings of the National Academy of Sciences}
  \textbf{\bibinfo{volume}{37}}, \bibinfo{pages}{760} (\bibinfo{year}{1951}).

\bibitem[{\citenamefont{Marshall et~al.}(1979)\citenamefont{Marshall, Olkin,
  and Arnold}}]{marshall1979inequalities}
\bibinfo{author}{\bibfnamefont{A.~W.} \bibnamefont{Marshall}},
  \bibinfo{author}{\bibfnamefont{I.}~\bibnamefont{Olkin}}, \bibnamefont{and}
  \bibinfo{author}{\bibfnamefont{B.~C.} \bibnamefont{Arnold}},
  \emph{\bibinfo{title}{Inequalities: theory of majorization and its
  applications}}, vol. \bibinfo{volume}{143} (\bibinfo{publisher}{Springer},
  \bibinfo{year}{1979}).

\bibitem[{\citenamefont{Rosenhaus}(2019)}]{rosenhaus2019introduction}
\bibinfo{author}{\bibfnamefont{V.}~\bibnamefont{Rosenhaus}},
  \bibinfo{journal}{Journal of Physics A: Mathematical and Theoretical}
  \textbf{\bibinfo{volume}{52}}, \bibinfo{pages}{323001}
  (\bibinfo{year}{2019}).

\bibitem[{\citenamefont{Garc{\'\i}a{-}Garc{\'\i}a and
  Verbaarschot}(2016)}]{GV16}
\bibinfo{author}{\bibfnamefont{A.}~\bibnamefont{Garc{\'\i}a{-}Garc{\'\i}a}}
  \bibnamefont{and}
  \bibinfo{author}{\bibfnamefont{J.}~\bibnamefont{Verbaarschot}},
  \bibinfo{journal}{Physical Review D} \textbf{\bibinfo{volume}{94}},
  \bibinfo{pages}{126010} (\bibinfo{year}{2016}).

\bibitem[{\citenamefont{Garc{\'\i}a{-}Garc{\'\i}a
  et~al.}(2018)\citenamefont{Garc{\'\i}a{-}Garc{\'\i}a, Jia, and
  Verbaarschot}}]{GJV18}
\bibinfo{author}{\bibfnamefont{A.}~\bibnamefont{Garc{\'\i}a{-}Garc{\'\i}a}},
  \bibinfo{author}{\bibfnamefont{Y.}~\bibnamefont{Jia}}, \bibnamefont{and}
  \bibinfo{author}{\bibfnamefont{J.}~\bibnamefont{Verbaarschot}},
  \bibinfo{journal}{Journal of High Energy Physics}
  \textbf{\bibinfo{volume}{2018}}, \bibinfo{pages}{1} (\bibinfo{year}{2018}).

\bibitem[{\citenamefont{Feng et~al.}(2019)\citenamefont{Feng, Tian, and
  Wei}}]{FTW19}
\bibinfo{author}{\bibfnamefont{R.}~\bibnamefont{Feng}},
  \bibinfo{author}{\bibfnamefont{G.}~\bibnamefont{Tian}}, \bibnamefont{and}
  \bibinfo{author}{\bibfnamefont{D.}~\bibnamefont{Wei}},
  \bibinfo{journal}{Peking Mathematical Journal} \textbf{\bibinfo{volume}{2}},
  \bibinfo{pages}{41} (\bibinfo{year}{2019}), ISSN \bibinfo{issn}{2096-6075},
  \urlprefix\url{https://doi.org/10.1007/s42543-018-0007-1}.

\bibitem[{\citenamefont{Feng et~al.}(2018)\citenamefont{Feng, Tian, and
  Wei}}]{FTW18}
\bibinfo{author}{\bibfnamefont{R.}~\bibnamefont{Feng}},
  \bibinfo{author}{\bibfnamefont{G.}~\bibnamefont{Tian}}, \bibnamefont{and}
  \bibinfo{author}{\bibfnamefont{D.}~\bibnamefont{Wei}}, \bibinfo{type}{Tech.
  Rep.} \bibinfo{number}{1806.05714}, \bibinfo{institution}{arXiv}
  (\bibinfo{year}{2018}).

\bibitem[{\citenamefont{Feng et~al.}(2020)\citenamefont{Feng, Tian, and
  Wei}}]{FTW20}
\bibinfo{author}{\bibfnamefont{R.}~\bibnamefont{Feng}},
  \bibinfo{author}{\bibfnamefont{G.}~\bibnamefont{Tian}}, \bibnamefont{and}
  \bibinfo{author}{\bibfnamefont{D.}~\bibnamefont{Wei}},
  \bibinfo{journal}{Random Matrices. Theory and Applications}
  \textbf{\bibinfo{volume}{9}}, \bibinfo{pages}{2050001, 24}
  (\bibinfo{year}{2020}), ISSN \bibinfo{issn}{2010-3263}.

\bibitem[{\citenamefont{Liu et~al.}(2018)\citenamefont{Liu, Chen, and
  Balents}}]{liu2018quantum}
\bibinfo{author}{\bibfnamefont{C.}~\bibnamefont{Liu}},
  \bibinfo{author}{\bibfnamefont{X.}~\bibnamefont{Chen}}, \bibnamefont{and}
  \bibinfo{author}{\bibfnamefont{L.}~\bibnamefont{Balents}},
  \bibinfo{journal}{Physical Review B} \textbf{\bibinfo{volume}{97}},
  \bibinfo{pages}{245126} (\bibinfo{year}{2018}).

\bibitem[{\citenamefont{Haldar et~al.}(2021)\citenamefont{Haldar, Tavakol, and
  Scaffidi}}]{haldar2021variational}
\bibinfo{author}{\bibfnamefont{A.}~\bibnamefont{Haldar}},
  \bibinfo{author}{\bibfnamefont{O.}~\bibnamefont{Tavakol}}, \bibnamefont{and}
  \bibinfo{author}{\bibfnamefont{T.}~\bibnamefont{Scaffidi}},
  \bibinfo{journal}{Physical Review Research} \textbf{\bibinfo{volume}{3}},
  \bibinfo{pages}{023020} (\bibinfo{year}{2021}).

\bibitem[{\citenamefont{Bravyi and Gosset}(2017)}]{bravyi2017complexity}
\bibinfo{author}{\bibfnamefont{S.}~\bibnamefont{Bravyi}} \bibnamefont{and}
  \bibinfo{author}{\bibfnamefont{D.}~\bibnamefont{Gosset}},
  \bibinfo{journal}{Communications in Mathematical Physics}
  \textbf{\bibinfo{volume}{356}}, \bibinfo{pages}{451} (\bibinfo{year}{2017}).

\bibitem[{\citenamefont{Boutin and Bauer}(2021)}]{boutin2021quantum}
\bibinfo{author}{\bibfnamefont{S.}~\bibnamefont{Boutin}} \bibnamefont{and}
  \bibinfo{author}{\bibfnamefont{B.}~\bibnamefont{Bauer}},
  \bibinfo{journal}{Physical Review Research} \textbf{\bibinfo{volume}{3}},
  \bibinfo{pages}{033188} (\bibinfo{year}{2021}).

\bibitem[{\citenamefont{Bravyi}(2004)}]{bravyi2004lagrangian}
\bibinfo{author}{\bibfnamefont{S.}~\bibnamefont{Bravyi}},
  \bibinfo{journal}{arXiv preprint quant-ph/0404180}  (\bibinfo{year}{2004}).

\end{thebibliography}
\end{document}